\documentclass[preprint, 12pt]{aastex}
\usepackage{natbib,afterpage}

\shortauthors{Soria et al.}
\shorttitle{Birth of an M83 ULX} 

\newcommand{\einstein}{{\it Einstein}}
\newcommand{\rosat}{{\it ROSAT}}
\newcommand{\chandra}{{\it Chandra}}
\newcommand{\xmm}{{\it XMM-Newton}}
\newcommand{\swift}{{\it Swift}}
\newcommand{\asca}{{\it ASCA}}

\newcommand{\hst}{{\it HST}}

\newcommand\LUM{\:{\rm ergs\:s^{-1}}}
\newcommand\FLUX{\:{\rm ergs\:cm^{-2}\:s^{-1}}}

\newcommand\eg{{\it e.g.}}

\begin{document}
\title{The Birth of an Ultra-Luminous X-ray Source in M83\footnote{
Based on observations made with NASA's Chandra X-ray Observatory, 
the NASA/ESA Hubble Space Telescope,
\swift\ , 
the 6.5 meter Magellan Telescopes located at Las Campanas Observatory, 
and the Gemini Observatory.
NASA's Chandra Observatory is operated by Smithsonian Astrophysical Observatory 
under contract \# NAS83060 and the data were obtained through program GO1-12115.
The \hst\ observations were obtained at the Space Telescope Science Institute,
which is operated by the Association of Universities for Research in Astronomy, Inc. (AURA),
under NASA contract NAS 5-26555.
The new \hst\ observations were obtained through programs \# GO-12513 and GO-12683. Data 
in the \hst\ archive from program GO-11360 was also used.
The ground-based observations were obtained through NOAO
which is operated by Association of Universities for Research in Astronomy, Inc. 
for the National Science Foundation.}}

\author{Roberto Soria\altaffilmark{2},
K. D. Kuntz\altaffilmark{3},
P. Frank Winkler\altaffilmark{4, 7},\\
William P. Blair\altaffilmark{3, 7},
Knox S. Long\altaffilmark{5, 7},
Paul P. Plucinsky\altaffilmark{6}, and 
Bradley C. Whitmore\altaffilmark{5}
}
\altaffiltext{2}{Curtin Institute of Radio Astronomy, Curtin University, 
1 Turner Avenue, Bentley WA6102, Australia; roberto.soria@icrar.org}
\altaffiltext{3}{The Henry A. Rowland Department of Physics and Astronomy, 
Johns Hopkins University, 3400 N. Charles Street, Baltimore, MD, 21218; 
kuntz@pha.jhu.edu, wpb@pha.jhu.edu}
\altaffiltext{4}{Department of Physics, Middlebury College, Middlebury, VT, 05753; 
winkler@middlebury.edu}
\altaffiltext{5}{Space Telescope Science Institute, 3700 San Martin Drive, 
Baltimore, MD, 21218;  whitmore@stsci.edu, long@stsci.edu}
\altaffiltext{6}{Harvard-Smithsonian Center for Astrophysics, 60 Garden Street, 
Cambridge, MA 02138; plucinsky@cfa.harvard.edu}
\altaffiltext{7}{Visiting Astronomer, Gemini Observatory, La Serena, Chile}


\begin{abstract}

A previously undetected ($L_X<10^{36}$ erg s$^{-1}$) source
in the strongly star-forming galaxy M83
entered an ultraluminous state between August 2009 and December 2010.
It was first seen with \chandra\ on 23 December 2010 at
${L_X \approx 4 \times10^{39}}$ erg s$^{-1}$, 
and has remained ultraluminous through 
our most recent observations in December 2011, 
with typical flux variation of a factor of two.
The spectrum is well fitted by a combination of absorbed power-law
and disk black-body models.
While the relative contributions of the models varies with time,
we have seen no evidence for a canonical state transition.
The luminosity and spectral properties are consistent 
with accretion powered by a black hole 
with $M_{BH} \approx 40$--$100 M_{\odot}$.
In July 2011 we found a luminous, blue optical counterpart 
which had not been seen in deep \hst\ observations obtained in August 2009.
These optical observations suggest that the donor star 
is a low-mass star undergoing Roche-lobe overflow, 
and that the blue optical emission seen during the outburst
is coming from an irradiated accretion disk.
This source shows that ultraluminous X-ray sources (ULXs)
with {\it low-mass} companions
are an important component of the ULX population in star-forming galaxies,
and provides further evidence that the blue optical counterparts
of some ULXs need not indicate a young, high-mass companion,
but rather that they may indicate X-ray reprocessing.
\end{abstract}

\section{Introduction} 

Point-like X-ray sources with luminosities exceeding the Eddington limit 
of normal stellar-mass black holes ($L_X \gtrsim 3 \times 10^{39} \LUM$, 
assuming isotropic emission) are commonly known as Ultraluminous  X-ray sources (ULXs).   
The most luminous class of non-nuclear sources in galaxies, 
they are widely believed to result from some extreme form 
of accreting X-ray binary containing a black hole.  
However, the nature of both the black holes that power the ULXs 
and the companion stars that fuel them remain enigmatic.  
Briefly, the principal competing models for the primaries are: 
(a) intermediate (between normal stars and AGN) mass black holes 
($ 10^2 M_{\odot} \lesssim  M_{\rm BH} \lesssim 10^4 M_{\odot}$) 
perhaps formed in the collapsed core of massive star clusters---many 
of which are present in M83 \citep{colbert99}; 
(b)  ``normal'' stellar-mass black holes ($M_{\rm BH} \la 20 M_{\odot}$) 
that are accreting well above their Eddington limit \citep{begelman02}
and/or that have  beamed emission \citep[\eg,][]{king01, begelman06};
and (c) ``heavy'' stellar black holes 
($M_{\rm BH} \approx 30$--$70 M_{\odot}$)---perhaps 
formed from direct collapse of metal-poor, massive stars 
\citep{belczynski10, zampieri09, pakull02}---that 
are accreting near or just above their Eddington limit.

A number of lines of evidence suggest that most ULXs 
are associated with young, high-mass stellar populations---the extreme 
of the high-mass X-ray binary (HMXB) population.
Statistically, ULXs are found mostly in star-forming spirals 
or irregular galaxies \citep{irwin04, swartz04}
and the number of ULXs per galaxy increases
with the star-formation rate \citep{colbert04, liu06}.
The cumulative luminosity function for ULXs 
in star-forming galaxies is consistent with the extrapolation 
of that for HMXBs \citep{walton11, swartz11}.
Within individual galaxies, 
$\sim$75\% of ULXs are found in thin spiral arms and dust lanes,
and tend to have high absorbing column densities \citep{liu06}.  
Theoretical considerations suggest that high-mass donors
can maintain a ULX in a persistent high state for up to 10 Myr 
\citep{rapetal05}.

Yet there is mounting evidence for a second population of ULXs, 
a population that represents an extreme form of low-mass X-ray binaries (LMXBs).  
The cumulative luminosity function of ULXs in elliptical galaxies
is consistent with the extrapolation of the luminosity function for LMXBs 
\citep{swartz04, walton11}, suggesting that those ULXs are extreme LMXBs.
In spiral galaxies, 
the number of ULXs is not purely a function of the star formation rate,
but is also a function of the total stellar mass,
suggesting that 15\% to 25\% of the total number of ULXs
are associated with the older stellar population
\citep{colbert04, liu06, winter06, mushotzky06}.

Point-like, blue optical counterparts have been  identified 
for a number of ULXs \citep[\eg,][]{ptak06, grise08}, 
thus reinforcing the link with a young, high-mass stellar population.   
However, some ULX counterparts that were initially identified
as massive, early-type stars on the basis of their blue colors 
\citep[e.g.][]{kuntz05}
are now thought to be low-mass stellar donors
and the blue colors due to optical emission from the accretion disk 
and/or reprocessed radiation from the X-ray source
\citep[e.g.][]{copperwheat07}.  
Indeed, \citet{tao11} have argued that the optical emission 
from the majority of ULX counterparts is dominated by X-ray reprocessing.
However, there have been no unequivocal identifications 
of low-mass donors in such systems, 
because we have never had a chance to observe them in quiescence.

In this paper we report the discovery of a ULX
that recently erupted in the nearby \citep[$d = 4.6\pm 0.2$ Mpc,][]{saha06} 
grand-design spiral galaxy M83 (NGC\,5236),  
and that we can confirm to be powered by a low-mass companion.   
The first exposures of a large \chandra\ project to study M83,  
taken on 23 and 25 December 2010, revealed an unexpected gift:  
a ULX---a non-nuclear source with a flux comparable 
to the total circumnuclear flux---that 
had not appeared in previous \chandra\ images from 2000--01, 
nor in any other prior X-ray image of M83.
The new source appeared $\sim1\arcmin$ east of the nucleus, well away 
from the spiral arms and from the many regions with active star formation.   
Since its discovery, we have monitored the source for a year
with \swift , and with \chandra\ in our ongoing program observations.
We have also obtained optical images of the field containing the ULX 
from the Gemini South telescope and from the {\it Hubble Space Telescope} (\hst ), 
and we find that a new unresolved source, 
not seen in previous \hst\ observations of the same field, 
has appeared coincident with the ULX.  

The remainder of this paper is organized as follows: 
In \S 2 we present the new \chandra\ and \swift\ data, 
along with a brief survey of  archival X-ray data where the ULX was absent, 
and in \S 3 we give the results of spectroscopic and timing analyses.  
In \S 4 we describe the optical observations and results.   
In \S 5 we argue that the M83 ULX is powered by  accretion from a low-mass companion 
that has recently expanded to overflow its Roche lobe, 
and that the present bright optical emission results 
from the reprocessing of X-rays in the accretion disk.
We go on to discuss the probable geometry and black hole mass in some detail, 
and close in \S 6 with a brief summary.   

\section{X-ray Observations}

The object was discovered using \chandra ,
whose spectacular angular resolution allowed the quick and definitive
determination that no such source appeared in previous X-ray images.
The source was discovered when its $L_{0.3-10}\sim4\times10^{39} \LUM$,
a ULX by most definitions of that class.
The \chandra\ spectrum is unequivocally that of a power-law-like source,
but there are many classes of sources with similar spectra,
so the nature of the source was not immediately clear.
We turned to \swift\ to follow the short-term evolution of the source,
and to archival data from other missions 
to determine what might have been there in the past. 

\subsection{{\it Chandra} Discovery }

The recent \chandra\ data were obtained 
as part of a detailed study of M83 
(Program 12620596; Long P.I.).
The data, totalling 729 ks, were obtained in ten observations, 
each longer than 50 ks.
The observations are clustered in December 2010, March 2011, and August 2011,
with a final observation in December 2011
(See Table~\ref{tab:obsid}.)
We used the back-illuminated S3 chip for maximum soft response, 
since most of the M83 disk fits within its $8\arcmin\times8\arcmin$ field. 
We carried out the observations in the ``very faint" mode 
for optimum background subtraction.   
We filtered and analysed the data with standard imaging 
and spectroscopic tools, such as {\it dmcopy}, {\it dmextract} 
and {\it specextract}, in the {\small CIAO} 
Version 4.3 \citep{fruscione06} data analysis system.

As shown in the upper right panel of Figure~\ref{fig:find_chart},
the ULX appears $\sim1\arcmin$ east of the nucleus of M83.
A count-weighted mean of the centroid positions over the first nine 
observations gives a position 
$${
{\rm R.A.}\; (2000) = 13^{\rm h} 37^{\rm m} 05^{\rm s}.135 \pm 0^{\rm s}.014 ,\ \
{\rm Dec.}\; (2000) = -29\degr 52\arcmin 07\farcs2 \pm 0\farcs3
}$$ 
(90\% confidence level).
The root-mean-square scatter of the source positions derived 
from each \chandra\ observation is ${\sigma \approx 0\farcs25}$.
We also determined a source position using only 
the five observations during March and April 2011, 
in which the aimpoint was closest ($\approx76$\arcsec) to the ULX, 
when the point spread function (PSF) would be most narrow and symmetric, 
with a 90\% encircled energy radius $\approx 0\farcs9$; 
we obtained the same result as from the total average, 
with a root-mean-square scatter ${\sigma \approx 0\farcs15}$.
We confirmed the accuracy of the \chandra\ position further by using 
two sources in the S3 chip with both X-ray and radio detections: 
SN1957D (located $\approx 2\farcm3$ north-east of the nucleus) 
and a background radio galaxy ($\approx 1\arcmin$ north-west of the nucleus). 
The mean X-ray positions of those two sources 
are within $\approx  0\farcs2$ of the radio positions, 
measured from our Australia Telescope Compact Array observations at 5 Ghz; 
the radio positions have themselves an uncertainty of $\approx0\farcs2$.  
We conclude that the mean \chandra\ coordinates can be offset 
by no more than $\approx  0\farcs3$ from the true position.

The two earlier observations of M83, taken with the \chandra\ ACIS-S in 
2000 and 2001, do not show the presence of the ULX 
(see the upper left panel of Figure~\ref{fig:find_chart}).
Applying the Bayesian method of \citet{kraft1991}
to the number of detected counts in the source and background regions,
we estimate the net count rate in 2000--2001 to be 
$< 1.0 \times 10^{-4}$ ct s$^{-1}$ at the 90\% confidence level.
This corresponds to an emitted luminosity $\la 1 \times 10^{36}$ erg s$^{-1}$ 
for any of the spectral parameter values
found in the 2010--2011 series of observations.
The newly erupted source had brightened by at least a factor of 3000.

\subsection{{\em Swift} Monitoring }

After the discovery of the ULX in the \chandra\ data,
we monitored it with several series of short ($\approx3$ ks) observations
with the \swift\ X-ray Telescope (XRT), 
to follow its short-term evolution.

The ULX is somewhat over $0\farcm5$ outside
the extended emission from the M83 bulge,
in a region free of bright, strongly structured diffuse emission,
but in a region where the diffuse emission is still significant,
as can be seen in the lower right panel of Figure~\ref{fig:find_chart}.
Thus, we had to choose the background region 
for photometry from the \swift\ XRT, 
which has a pixel size of $2\farcs4$, carefully.
We set the source region to have a radius of $18\arcsec$,
which contains 70\% of the encircled energy at 1 keV,
and the background region to be an annulus 
stretching from $18\arcsec$ to $35\arcsec$,
avoiding several nearby regions of enhanced diffuse emission.
The background region contains the extended wings of the ULX,
comprising some 14\% of the total source flux\footnote{
The XRT instrument handbook provides formulae for calculating
the encircled energy at $0.5$ keV and $4.0$ keV.
We used these formula to calculate the encircled energy fractions
for $0.3$--$2.0$ keV (using the values for $0.5$ keV)
and for $2.0$--$10.0$ keV (using the values for $4.0$ keV).}.
Examination of the \chandra\ image for the source and background regions
show that both contain several faint point sources;
none of these is readily detectable in the \swift\ image,
suggesting that the error produced by not excluding them
is on the order of the Poisson statistics of the background.
We extracted counts from the source and background regions,
solving the simultaneous equations to get the total source counts
and the background counts per pixel. 
Since the exposures were relatively short,
the background rate was somewhat uncertain,
particularly in bands narrower than the full $0.3$--$10.0$ keV range of the detector.
In order to reduce this uncertainty, 
we determined the mean background rate over all the observations
and recalculated the source rates.   
The difference among individual background rates was small, 
and the signal-to-noise of the source counts 
was improved by using the time-averaged background.
The resultant light-curve is shown in Fig.~\ref{fig:lc};
the count rate varies from $\approx0.03$ to $\approx0.07$ count s$^{-1}$.

The $(2.0$--$10.0$ keV$)/(0.3$--$2.0$ keV) hardness ratios, 
shown in Figure~\ref{fig:cr_hr},
vary significantly among the observations,
making the use of a single energy-conversion factor inappropriate.
Instead, we calculated the fluxes and hardness ratios 
expected as a function of power-law index, 
assuming an absorbed power law with $N_H = 1.2\times10^{21}\: {\rm cm}^{-2}$,
determined from the best-fitting parameters 
for the December and March \chandra\ data.
Then, we used the measured count rates and hardness ratios to produce a flux for 
each \swift\ observation.
The \swift\ light curve in flux units is shown in 
the right-hand panel of Fig.~\ref{fig:lc}.
Even with 3 ks exposures, 
the total number of counts is insufficient for spectral fitting,
and the variation in the hardness ratio argues 
against summing different observations.
We obtained three \swift\ exposures 
that were nearly simultaneous with \chandra\ exposures.
The fluxes calculated from the \swift\ count rates and hardness ratios
are roughly consistent with those derived by fitting the \chandra\ spectra, 
after the latter were corrected for pile-up.

The only earlier \swift\ XRT observation useful for this work
was made in 2005.
It provides an upper limit of $F_X$(0.3\,-\,10 keV) $< 1.7 \times 10^{-14} \FLUX $
($L_X$(0.3\,-\,10 keV) $< 4.5 \times 10^{37} \LUM$),
assuming an absorbed power law
with $N_H = 1.2\times10^{21}\: {\rm cm}^{-2}$ and $\Gamma=2.0$.

\subsection{Previous X-ray Observations from Other Missions}

M83 has been a popular target in surveys of normal galaxies.
As a result there are sufficient data 
to determine whether the ULX in M83 has been bright in the past.

{\it XMM-Newton: }
Three observations of M83 are available in the public archive.
The galaxy is well centered in the field of view of the first of these, 
from 2003, while in the following two, from 2008, 
the galaxy falls on a peripheral chip,
so the ULX is not covered by all of the instruments 
in the last two observations.
We measured upper limits to the EPIC count rates 
using a source region with $r<18\arcsec$ 
and a background region with $18\arcsec<r<35\farcs$; 
we converted count rates to fluxes assuming an absorbed power law 
with $N_H = 1.2\times10^{21}\: {\rm cm}^{-2}$ and $\Gamma=2.0$.
If we assume that the source was constant over all the \xmm\ observations, 
the upper limit becomes $F_X (0.3$--$10.0$ keV) $< 3.9 \times 10^{-15} \FLUX $
($L_X (0.3$--$10.0$ keV) $< 1.0 \times 10^{37} \LUM$),
while the best individual instrument/exposure result
is for the PN in 2003 of $F_X (0.3$--$10.0$ keV) $< 7.3 \times 10^{-15} \FLUX $
($L_X (0.3$--$10.0$ keV) $< 1.8 \times 10^{37} \LUM$).

{\it ROSAT: }
There is one PSPC observation from 1993 
and two HRI observations from 1993 and 1994 in the archive,
each with exposure time $\approx24$ ks \citep{immler99}.  
There is no obvious source at the location of the ULX in the HRI exposures;
the detection limit for the combined HRI exposures is 
$F_X (0.1$--$2.4$ keV) $\approx10^{-14} \FLUX$
($L_X (0.1$--$2.4$ keV) $\approx2.5\times10^{37} \LUM$).
The nucleus and the ULX are not resolved in the PSPC image.
However, since in its ultraluminous state it is comparable to the nucleus in brightness,
had the source been ultraluminous during the PSPC observation,
one would expect the central source to have had an east-west elongation,
which was not observed.
Thus we may be reasonably confident that the source was not ultraluminous
during the PSPC observation.
The same argument can be applied to the \asca\ observation in February 1994,
and the \einstein\ IPC observation in July 1979.

{\it Einstein:}
Besides the IPC observation from 1979, 
with an exposure time of 6 ks, 
there are two HRI observations from 1980 and 1981 
with exposures of 25 and 20 ks, respectively \citep{trinchieri85}.  
Taking the dimmest detected source in the co-added HRI images from \citet{trinchieri85}
as an extreme upper limit, we find that the flux is $F_X<1.3\times10^{-13} \FLUX$
($L_X < 3.3\times10^{38} \LUM$).

{\it Summary:}  
M83 was observed to a depth that would have revealed a ULX,
or even a moderately bright ($L_X>10^{37}$ erg s$^{-1}$) X-ray binary
in 1979, 1980, 1981, 1993, 1994, 2000, 2001, 2003, 2005, and 2008.
The continuously high emission for at least 12 months, 
from late December 2010 through late December 2011, suggests that 
if the source has repeated outbursts, they are relatively long.
Thus the non-detections over the last three decades suggest 
that this source is either new or has a long period between ultraluminous episodes.

\section{X-ray Results}

The \chandra\ data are magnificent, 
providing spectra with $>10^4$ counts, 
but have sparse temporal coverage.
The \swift/XRT data provide a more complete light-curve,
but can provide no more than a hardness ratio.
The analysis of each informs and is informed by the other.

\subsection{Temporal Variation}

The \swift\ count rate light-curve in Figure~\ref{fig:lc}
shows that the count rate varied within a range of roughly a factor of two.
The data hint at a flux decline since March 2011, 
which is in agreement with the \chandra\ flux trend. 
However, given that the source showed flux variations 
of comparable amplitude during January and February, 
it may be too early to conclude that the outburst is near its end.
The spectral shape,
as tracked by the $2.0$--$10.0$ keV$/0.3$--$2.0$ keV hardness ratio,
is also strongly variable.
The count rate versus hardness ratio plot in Figure~\ref{fig:cr_hr}
suggests two different regimes of behavior.
In epochs 2 to 5, as defined in Table~\ref{tab:obsid}, 
the hardness ratio varied strongly 
while the count rate did not, 
staying within $\approx0.03$ to $0.045$ count s$^{-1}$.
In epochs 6 to 13 the hardness ratio did not vary significantly,
while the count rate did.
After epoch 13 the source returned to the mode seen in the earlier epochs.
The broad-band count rate and the hardness ratio are not well-correlated,
and do not follow a characteristic track in the hardness-intensity diagram.

The broad-band flux light-curve derived 
from a combination of \swift\ and \chandra\ observations
also fluctuates within a range of roughly a factor of two,
between 1 and 2$\times10^{-12} \FLUX$.
We determined the \chandra\ fluxes from multicomponent spectral fits.
We derived the \swift\ fluxes by assuming a power-law spectrum 
absorbed by a column of $1.2\times10^{21}$ cm$^{-2}$,
constraining the photon index from the hardness ratio,
and normalizing the flux to obtain the measured
$0.3$--$10.0$ keV band count rate.
The full range of flux was spanned 
by observations in epochs 2 and 3 (a day apart)
and by epochs 1 and 2 (9 days apart) and 6 and 7 (11 days apart).
There is a hint of greater variability in the observations
before mid-February (epoch 8) than after;
the clump of observations in March (epochs 9-11) 
shows little variation in count rate or flux.
Unlike the count rate {\it versus} hardness ratio plot,
the flux {\it versus} hardness ratio plot shows little structure,
and the conversion from \swift\ count rates to fluxes adds scatter to the data.

Overall, the source variability during 
our first 12 months of observations is more similar to the variability 
seen in ULXs such as Holmberg II X-1 \citep{grise08} and Holmberg IX X-1 
\citep{kaaret09}, than to canonical state transitions, routinely seen 
in Galactic black hole binaries \citep{fender04, mcclintock06}.

We also investigated the intra-observational variability 
for each \chandra\ epoch. 
We extracted background-subtracted lightcurves 
with the {\small CIAO} task {\it dmextract}, and analysed them 
using the {\it lcstats}, {\it efsearch} 
and {\it powspec} tasks in {\small FTOOLS} \citep{blackburn95}.
We computed the power spectrum from the shortest period
not biased by the Chandra readout rate ($\sim10$ s) 
to the length of the shortest observation ($\sim50$ ks).
We do not find significant features in the power spectral density 
over this $0.1$ to $(2 \times 10^{-5})$ Hz band for any observation,
nor do we see any dips, eclipses or flares.
The $\chi^2$ probability of a constant light-curve 
is $\approx 1$ for each epoch;
the Kolmogorov-Smirnoff probability \citep{kolmogorov41} 
that the flux is constant is $\ga 7\%$
for all observations except for that of September 2011
(when it is $\approx 4 \times 10^{-3}$). 
We can only place a $3\sigma$ upper limit of $\approx 45\%$ ($\approx 16\%$) 
to the RMS fractional variation for 10 s (100s) bins at any epoch.
Overall, the short-term variability is unremarkable. 
\citet{heil09} has demonstrated that ULXs show a wide variety
of short-term variability,
including many that show no signs of variability.

\subsection{X-ray Spectroscopy}

For each \chandra\ observation,
we extracted source spectra from a circular region with a $4\arcsec$ radius,
and background spectra from an annular region between radii 
of $6\arcsec$ and $12\arcsec$; we verified that the background region
contains no other point sources 
and only a negligible amount of diffuse emission.
We used the {\it specextract} script 
in {\small CIAO} Version 4.3 \citep{fruscione06}
to build response and area-corrected ancillary response files,
and we fitted the spectra in {\small XSPEC} Version 12 \citep{arnaud96}.
Two fully representative examples of the \chandra\ spectra 
and the fits we carried out are shown in Figure~\ref{fig:chandra_spec}.

We fitted the $0.3$--$10$ keV spectrum 
from each of the \chandra\ observations 
with two {\small XSPEC} models commonly applied 
to the study of black-hole X-ray binaries: 
an absorbed disk blackbody ({\it diskbb}) plus power-law model, 
and an absorbed thermal Comptonization ({\it comptt}) model \citep{titarchuk94}.  
The {\it diskbb} plus {\it power-law} model in ULXs 
is typically a good but purely phenomenological test 
for the presence of a soft excess below 1 keV.
Under certain assumptions, 
the soft excess emission may be attributed to the accretion disk 
and may be used to constrain the inner disk size.
The {\it comptt} model is a suitable test for mild spectral curvature,
in particular at the high energy end of the ACIS bandpass; 
a characteristic class of ULX spectra with a downturn around 5 keV 
is formally well fitted with Comptonization models at low plasma temperatures, 
$kT_e\approx1.5$ to $2$ keV \citep{stobbart06,roberts07,gladstone09}.

The Galactic absorbing foreground column is 
$4\times10^{20}$ cm$^{-2}$ \citep{lab}
while the column density due to M83
in this direction is $1.5\times10^{20}$ cm$^{-2}$
\citep[using the naturally weighted map from][]{things}.
Thus, other than any internal absorption,
the emission should be only lightly absorbed by intervening components.
As a result of these considerations, 
we assumed a fixed Galactic foreground absorption 
with a column density of $4\times10^{20}$ cm$^{-2}$,
and a variable absorption from within M83 and the system itself.
These column densities are roughly comparable to the optical extinction 
(see Section 4.2). 
 
At the observed ACIS-S count rate $\approx0.2$ ct s$^{-1}$
(Tables~\ref{tab:fit-pldbb} and~\ref{tab:fit-comptt}),
the source spectra are affected by pile-up\footnote{
http://cxc.harvard.edu/ciao/download/doc/pileup\_abc.pdf}, 
meaning that there is a high probability that more than one photon falls 
in a single pixel (or adjacent pixels) between frame readouts.
In general, one can correct for the spectral distortions caused 
by pile-up either by removing pixels with a high pile-up probability 
from the spectral extraction regions, 
or by using convolution models during spectral fitting.
We started with the latter technique, and used 
the {\it pileup} convolution model of \citet{davis2001} within {\small XSPEC}.
This model has two principal parameters which can be allowed to vary: 
the grade migration parameter, $\alpha$,
and the fraction of the events in the source extraction region
to which the pileup will be applied, {\it psffrac}. 
When we allowed both of these parameters to vary, 
we usually found that there were two separate regions 
of parameter space that fit the \chandra\ spectra.  
The two model fits implied very different pile-up corrected count rates 
and different ranges of the $\alpha$ and the {\it psffrac} parameters, 
but otherwise produced the same spectral shape 
and goodness of fit, as determined by $\chi^2$.
To break the degeneracy and select the proper range for $\alpha$ and {\it psffrac},
we extracted another set of source spectra from annuli 
that excluded the piled-up central pixels.
We then constrained the $\alpha$ and {\it psffrac} parameters
so that they would produce a spectral fit to the piled-up data
that was consistent with the fits to the spectra extracted from the annuli. 
That this rather involved process produces the correct flux
is supported by the several nearly simultaneous \swift\ observations
for which the pile-up corrected \chandra\ fluxes are roughly consistent 
with their \swift\ counterparts.

We find that a {\it diskbb} component is significantly detected
in five of the observations but not the other five (Table~\ref{tab:fit-pldbb}).
Although a cool disk-blackbody component produces a marginal improvement 
in the $\chi^2$ for two of those four observations,
the F test shows that it is not statistically significant 
(F-test probability $P = 0.654$ for the 23 December 2010 observation, 
$P = 0.203$ for 4 September 2011). 
For the 28 December 2011 observation the disk-blackbody produces
a significant improvement in the fit (F-test probability $P = 0.011$)
although the disk-blackbody component itself
is only marginally detected (see Table~\ref{tab:fit-pldbb}).
We see no evidence for a disk-blackbody component
in the remaining two observations.
A soft excess is detected significantly in the 3 April 2011 observation
(F-test probability $P = 1.1 \times 10^{-3}$), 
but at very low temperatures ($kT < 0.16$ keV)
in an energy range where the ACIS-S3 sensitivity is low
and most of the emitted flux is absorbed. 
Thus, we can neither determine whether that soft excess is a {\it diskbb} component, 
nor reliably constrain its characteristic temperature and emission radius. 

The absence of a {\it diskbb} component in some observations 
cannot be due simply to shorter exposure times. 
Had such a component always been as strong as during the March 2011 observations, 
when it accounted for almost half of the emitted X-ray flux, 
it would have been easily detected at every other epoch. 
It is implausible that an accretion disk would form,
for example, in the two days between 23 and 25 December 2010, 
or vanish between 29 March and 3 April 2011. 
Thus, we suggest that the differing significance of the {\it diskbb} component 
in various observations more likely
depends on the fraction of the inner-disk photons that are upscattered
in a rapidly variable Comptonizing medium; 
it is this upscattering that is probably responsible for the power-law component. 

As the observations from March 2011
are those where the {\it diskbb} component is most prominent, 
they give us the best direct view of the inner disk, 
and allow us to use those observations to constrain the disk parameters.
The characteristic peak temperatures at those epochs are $\approx 0.3$ to $0.4$ keV
corresponding to a characteristic inner radius 
$r_{\rm in}/(\cos \theta)^{1/2} \approx 700$ to $1000$ km.
If $r_{\rm in}$ corresponds to the innermost stable circular orbit 
of a face-on Schwarzschild black hole, 
this value would suggest a black hole mass $\approx 80$ to $100 M_{\odot}$,
or higher for a fast-spinning black hole.
On the other hand, 
the classical Eddington-luminosity argument provides a lower limit of
$\approx 40 M_{\odot}$ for the black hole mass for the same data,
L$_X (0.3-10. keV)\sim5\times10^{39}$ ergs s$^{-1}$.
These two estimates stand in apparent contrast to one another. 
However, we have no reason to assume that we are seeing disk emission 
from near the innermost stable circular orbit.
If the inner part of the disk is covered by an optically-thick scattering region,
the thermal component should come from a larger radius. 
This is likely to be the case if the mass accretion rate is super-critical, 
that is above the threshold at which a radiatively-efficient,
geometrically-thin disk cannot survive. 
At such accretion rates (which may be the defining characteristic of ULXs), 
the inner region may be shrouded by a mass-loaded, optically-thick outflow, 
and thus the observed temperature of the soft thermal component
would be the temperature at the photosphere of such outflow, 
while the fitted inner-disk radius would be the radius 
at which the disk begins to be covered or replaced by the outflow
\citep[e.g.][]{kp2003, plfba2007, soria07, gladstone09}.
It has been suggested \citep{gladstone09} that a ULX with a lower diskbb temperature
($kT_{\rm in} \la 0.2$ keV) and relatively weak soft thermal component
compared to the hard power-law component
will have a stronger outflow and greater mass loading,
and thus a lower electron temperature in the wind-dominated,
Comptonizing inner region (``warm corona'' at $kT_e \approx 2$ keV), 
which results in a spectral break at energies $\sim 5$ keV.

In our case, the power-law component is dominant 
or comparable to the {\it diskbb} emission at all epochs. 
The strong power-law component suggests that the inner disk 
may indeed remain partly shrouded even in the March 2011 observations, 
and hence the characteristic masses and sizes based on $r_{\rm in}$ 
must be taken as upper limits. 
Fitting the spectra with the {\it comptt} model helps us investigate this issue. 
For all observations,
the {\it comptt} model provides statistically equivalent fits
to the {\it diskbb} plus {\it power-law} model.
For all except the 23 December 2010 observation, 
we can place only a lower limit (typically, $T_e \ga 2$ keV)
to the best-fitting electron temperature in the scattering region;
this is another way of saying that the power-law-like portion of the spectrum
extends beyond the upper limit of the ACIS energy band, 
without a sharp downturn at least up to $\approx7$ keV.
To test for the presence of a high-energy spectral break,
we fitted the $2.0$--$10$ keV portion of the spectrum 
with both power-law and broken power-law models, 
after fixing the pile-up model parameters.
We find (Table~\ref{tab:fit-pow}) 
that a broken power-law model does not improve the fit 
in any of the \chandra\ observations.
The lack of a break in the power-law suggests either 
that this ULX was not in a slim-disk/optically-thick warm-corona state \citep{roberts07, gladstone09}, 
or that its characteristic coronal temperature was higher 
than typically found in that variety of ULXs. 
We conclude that the accretion rate was only moderately super-critical, 
perhaps not high enough to launch a massive outflow, 
and that the spectral appearance of this ULX is closer to 
the ``very high state'' of Galactic BH transients 
than to the most extreme examples of warm-corona/outflow sources 
described in \citet{gladstone09}.

With only a few exceptions,
both the {\it powerlaw}$+${\it diskbb} fits and the {\it comptt} fits
indicate that the source has very little internal absorption.
The {\it diskbb} fits suggest intrinsic absorbing column densities 
$\sim$ a few times $10^{20}$ cm$^{-2}$,
somewhat larger than the local \ion{H}{1} column, $1.5\times10^{20}$ cm$^{-2}$;
the {\it comptt} fits suggest values consistent with no M83 absorption at all.

Finally, we searched for emission lines, 
either from the ULX itself ({\it e.g.}, Fe K lines) 
or from spatially unresolved, X-ray ionized gas in its surroundings. 
We combined the spectra and their respective response files 
from all the 2010--2011 {\it Chandra} observations
to increase the signal-to-noise ratio of any possible features. 
We do not find any significant lines or edges. 
A narrow 6.4 keV emission line could be added
to the local best-fitting continuum,
but can have an equivalent width no greater than $\sim12$ eV (90\% confidence level). 
Non-detection of either broad or narrow fluorescent Fe lines is not
a surprising feature for a ULX \citep[e.g.,][]{goad06}, although
such lines have been detected in the peculiar case of a ULX in M82
\citep{strickland07, dewangan06}.
In general, Fe lines from stellar-mass black holes are weaker than those
from AGN \citep{ross07}, because iron in the inner region
of the accretion disk is more highly ionized, and the blurring
of the line profile due to Compton scattering is more severe.
This is the case in particular when the power-law-like X-ray
emission from a scattering corona is comparable to or stronger than
the direct thermal emission from the accretion disk---as is the case
for most ULXs above 1 keV. 
If the inner region of the disk is covered or replaced 
by an optically thick scattering corona \citep{roberts07}, 
we do not expect to see any reflection features.
\citet{caballero10} proposed that ULX spectra
with a steepening around $\approx 5$--$7$ keV are dominated by
a reflection component with a relativistically broadened Fe line;
however, this interpretation remains somewhat controversial
as it requires {\it ad hoc} parameters. 
In any case, the continuum spectrum of the ULX in M83 
does not have such high-energy steepening, either.

In conclusion, had we not had previous observations of M83,
the X-ray properties of this source would look like 
those of any number of ULXs.
Its power-law photon index $\approx 2$, and its relatively low 
disk temperature are consistent with average ULX properties 
\citep{feng11,berghea08}.
Flux variability by a factor of a few on timescales of days (and longer)
is typical of ULXs \citep{grise08,kaaret09}, 
though statistics on that variabilty are sketchy, 
and the most closely monitored sources are brighter than this one. 
Finally, the lack of a disk-dominated phase 
of the outburst (known as high/soft state in Galactic black holes) 
is also a typical ULX property \citep{soria11}.

\section{Optical Counterpart}

Optical counterparts of persistent ULXs are typically faint,
usually with a B-type star-like appearance at extragalactic distances. 
When significantly detected, optical variability is only a fraction 
of a magnitude \citep[e.g.,][]{zampieri11} 
and its physical interpretation is unclear.
But in this case, given the proximity of M83 
and the dramatic change in the X-ray flux,
we have a unique opportunity to investigate and understand 
the corresponding changes in the optical and UV fluxes.
The change was, as expected, below the detectability of the \swift\ UVOT,
but was recorded with high confidence by \hst\ and ground-based observatories.

\subsection{Ground-Based Observations}

As part of a previously planned Gemini program,
we were able to image the section of M83 containing the ULX
from the 8.2-m Gemini-South telescope on 8 April 2011 (UT)
(Program \# 2011A-0436, Winkler PI).  
We used the Gemini Multi-Object Spectrograph (GMOS) in its imaging mode, 
which has a field 5\farcm5 square and 0\farcs0728 pixels. 
We binned the image $2 \times 2$ for an effective pixel size of $0\farcs 146$,  
appropriate for the seeing of $\approx0\farcs 7$ at the time of our observations.  
Four filters were used: $u^\prime$, $g^\prime$, $r^\prime$, and $i^\prime$, 
with individual exposures of 600, 100, 150, and 200 s, respectively, 
and with four dithered exposures through each filter.  
We processed the image data using standard IRAF\footnote{
IRAF is distributed by the National Optical Astronomy Observatories, 
which is operated by the  AURA, Inc., under cooperative 
agreement with the National Science Foundation.}   
and Gemini reduction procedures for overscan and bias subtraction, 
flat-fielding, aligning, and stacking.   
Flux calibration was done using observations of the spectrophotometric
standard LTT 4316 \citep[see][]{hamuy92} taken immediately after the M83
images.  All observations were at airmass $< 1.1$.

A blue stellar object, faint but clearly visible,
is present at the position of the ULX in the stacked images 
in the $u^\prime$, $g^\prime$, and $r^\prime$ bands, 
and possibly in $i^\prime$ as well (Figure~\ref{fig:gem_opt_3}, central panel). 
Even though the source is located in an inter-arm region 
well within the innermost spiral arm of M83, 
the background from the galaxy is complex enough 
to make accurate photometry of a faint object difficult.   

We had previously taken images of M83 with the IMACS instrument 
on the 6.5-m Magellan I telescope in April 2009 
as part of a program to inventory supernova remnants and other nebulae in M83
(Program \# 2009A-0446, Winkler PI).   
These images were taken through narrow-band filters 
and so are not directly comparable to the 2011 Gemini images, 
but they nevertheless show fainter stars than the 2011 images 
because of long exposures and exceptional seeing, $\approx0\farcs4$.    
No object at the ULX position is visible in these 2009 images 
with an upper limit $m_{5150} \gtrsim 25$ mag.
Although the bands are far from identical, 
we can use the Magellan images 
(after PSF matching and scaling)
to subtract much of the stellar contribution in the Gemini images near the ULX, 
to produce a background that is far more uniform 
and thus more amenable to aperture photometry.   
Figure~\ref{fig:gem_opt_3} 
shows both a Gemini and a Magellan image of the region, 
together with a difference image that clearly shows the ULX counterpart.
From the $u^\prime$ and $g^\prime$ images 
we subtracted the sum of [O\,III] plus a 200 \AA\ wide 
continuum band centered at 5200 \AA; 
from the $r^\prime$ image we subtracted one 
in a 150 \AA\ wide continuum band centered at 6800 \AA.  
We used these difference images for conventional aperture photometry 
of the ULX counterpart to obtain 
the Gemini AB magnitudes given in Table~\ref{tab:opti}. 
The uncertainties quoted include any due to possible residuals 
from imperfectly subtracted stars and diffuse background.

\subsection{{\em Hubble} Observations}

Images with better angular resolution and higher sensitivity 
for the M83 field containing the ULX 
were taken in August 2009 with the newly installed WFC3 instrument on \hst\ 
as part of the Early Release Science program \citep{dopm83}.
Within the error circle of the ULX position, there are no blue sources; 
the only objects are a few very faint red stars.   
Following the discovery of the ULX counterpart in the Gemini images, 
we requested and were awarded two orbits of Director's Discretionary \hst\ time 
for a second-epoch WFC3 observation, which was carried out on 27 July 2011
(Program 12683, PI: Winkler).
The data were processed with the standard WFC3 pipeline, 
including drizzling of the three dithered frames taken through each of the four filters.  
The ULX counterpart is readily apparent in the 2011 WFC3 images 
through the F336W ($U$), F438W ($B$), and F555W($V$) filters.   
Alignment and subtraction of the 2009 $I$ image 
renders the counterpart easily visible on the 2011 F814W ($I$) image as well.
The world coordinate systems associated with the WFC3 images 
at both epochs have been updated to agree 
with that which we determined for the Magellan images, 
which in turn was based on several hundred astrometric stars.
The position of the ULX counterpart is 
$${\rm R.A.}\; (2000.) = 13^{\rm h} 37^{\rm m} 05.127^{\rm s},  \ \ 
{\rm Decl.}\; (2000.) = -29\degr 52\arcmin 06.92\arcsec\;.$$
We estimate the absolute uncertainty in the coordinates to be, 
at worst, $0\farcs1$, 
and the registration of all the images relative to one another is better than 10 mas.  
The difference between the optical and X-ray coordinates is $0\farcs28$,
which is well within the combined uncertainty of the X-ray and optical positions.
WFC3 images from both the 2009 and the 2011 epochs are shown 
for a small region directly around the ULX position in Figure~\ref{fig:wfc3}.   

We have carried out photometry of the ULX counterpart 
and neighboring stars in the WFC3 images using DAOPHOT, 
consistently taking an aperture with a radius of 2 pixels (0\farcs08) 
and using standard encircled-energy corrections 
to account for the missing flux. 
In the 2009 images, there are a number of very faint red stars 
quite near the position of the ULX counterpart, 
though it is not clear if any of them is actually the donor star in quiescence.   
After updating the world coordinate systems of both images 
so they are accurately registered, 
we carried out photometry of the 2009 images 
using the identical positions and apertures as in 2011 
to obtain the limits given in Table~\ref{tab:opti}.    
While we used standard procedures for the July 2011 observations,
we used ``forced photometry'' at the position of the ULX for the July 2009 data.
We did see an excess at the position of the ULX,
and these are the values reported in Table~\ref{tab:opti}.
However, since there was no specific counterpart at this position,
the values represent upper limits to the flux from the counterpart. 

As part of an ongoing analysis of the HST data for M83, Hwihyun Kim
(private communication) provided photometry and reddening estimates
for stars surrounding the ULX position.  She determined a mean
total extinction of A$_V=0.30\pm0.34$.  This value indicates modest
but variable extinction, consistent with the appearance of the
HST imagery.  We adopt this mean value for the ULX in correcting
the values in Table~\ref{tab:opti} to obtain intrinsic fluxes. 
From the WFC3 data 
we find that the counterpart's absolute magnitude is $M_V = -4.85 $, 
and a total optical luminosity of 
$L(3300-9000 {\rm \AA}) \approx 2 \times 10^{37} \LUM$
at the time of the July 2011 observation.  
In quiescence, the counterpart was no brighter than $M_V \approx - 2.1 $, 
and therefore the optical counterpart 
has brightened optically by a factor of at least 10.  

Comparison between the AB magnitudes obtained on 8 April from Gemini
and those obtained on 27 July from \hst\ (Table~\ref{tab:opti}) indicates
no evidence for optical variability over this period.
Further monitoring of the optical source would be valuable,
either in tight coordination with X-ray observations,
or especially if the X-ray object starts to fade significantly.

\section{Discussion}

We have discovered a bright new source 
in the low-extinction inter-arm region of M83.
Although it is well within the D$_{25}$ radius,
it is still worth considering whether the source could be a background AGN.
Using a standard definition of the X-ray/optical flux ratio
$$\log{(f_X/f_{\rm R})} = \log{f_X}+5.5+m_{\rm R}/2.5$$ \citep{hornschemeier01}
where $f_X$ is the $0.3$--$10$ keV flux and $m_{\rm R}$ is the Cousins magnitude,
we find for this source, with $m_{\rm R}\approx24$ and $\log{f_X}\approx-12$, 
$\log{(f_X/f_{\rm R})}\approx3.5$.
This is typical of stellar-mass systems,
while AGN have typical $-1\la\log{(f_X/f_{\rm R})}\la1$ \citep{bauer04,laird09}.
Furthermore, AGN at this flux level are relatively rare
\citep[only $\approx0.1$ per square degree:][]{cappelluti09}
and nearly always have identified optical counterparts.
Finally, we are  not aware of any AGN that has been observed 
to vary in X-rays by the factor of at least 3000, 
as seen for the X-ray source in M83.
Thus, it is highly improbable that the source is a background AGN. 
There are no known historical supernovae in this region,
so it is unlikely that the sudden X-ray/optical increase is due to
a previously undetected remnant beginning to interact 
with the circum-stellar medium.

Our discovery of a transient ULX in an interarm region of M\,83 suggests
that the ULX population is more diverse than often assumed.
Most ULXs in nearby galaxies are variable sources 
but have been persistently active throughout the years since their original discovery  
(typically, with {\it ROSAT} in the 1990s). 
Instead, this source was not detected by \einstein , \rosat , \xmm , 
nor in previous \chandra\ observations. 
A flux increase of $> 3000$ between the 2000--2001 and the 2010--2011 
\chandra\ observations is very unusual for a ULX, 
but it resembles typical behavior of Galactic black hole transients.
Its current bright state has lasted at least twelve months, 
but likely less than two years 
(based on the faintness of its optical counterpart 
in the August 2009 {\it HST} observations).
This is longer than most Galactic black hole transients 
\citep[typically, a few weeks:][]{mcclintock06}, but is not unprecedented. 
For example, the 1996--1997 outburst of GRO J1655$-$40 
lasted for 15 months \citep{soria00}, 
and GRS 1915$+$105 has remained bright since 1992 \citep{castro-tirado11}.

The other defining characteristic of this ULX is that it is located 
far from any star-forming region, 
and it must have a low-mass, evolved donor star (mass $< 4 M_{\odot}$), 
since no OB counterparts were detected at the ULX position 
before the start of the outburst (see \S 5.2). 
\citet{irwin03, irwin04} argued that ULX candidates 
in the old stellar populations of elliptical galaxies
were mostly due to background AGN contamination. 
In spiral galaxies with mixed populations,
ULXs with low mass counterparts had been identified 
only in a statistical sense \citep{swartz04, walton11}.
However, our secure identification of this ULX with its optical counterpart
confirms the existence of two different classes of ULXs.
More specifically, 
this source is a ULX powered by accretion from a low-mass donor in an older environment
within a galaxy with active star formation. 
This suggests that a classification 
of ULXs based on the global properties of their host galaxies is incomplete,
as it may miss or underestimate a population of older, 
transient ULXs with short active phases,
particularly at the lower reaches of the ``ultra-luminous'' luminosity range.
The presence of older ULXs in star-forming galaxies
has long been proposed by \citet{mushotzky06}.
Further evidence for the existence of ULXs with low-mass donors 
in old stellar populations 
comes from two transient sources that are only a factor of two less luminous 
than the new ULX in M83: one in NGC\,5128 \citep{sivakoff08}
and one in M31 \citep{kaur11}. 

\subsection{The Nature of the Accretor} 

Beginning with \citet{shakura73},
a number of steady-state mechanisms have been proposed 
to allow large super-Eddington accretion rates
that lead to much milder super-Eddington luminosities.
Galactic black-hole binaries however, 
with a few exceptions \citep{jonker04},
stubbornly remain below the Eddington limit.
While it is accepted that ULXs are powered by accreting black holes, 
the main unsolved issue is whether they contain a different 
(more massive) kind of black hole than typical Galactic sources, 
or are simply in a different accretion state ({\it e.g.,} 
at a super-Eddington accretion rate).
We found that the $0.3$--$10$ keV spectrum of this source is typical of ULXs: 
dominated by a power-law with a photon index $\approx2$ 
(intermediate between the soft and hard state of Galactic black holes), 
with an additional soft thermal component at $kT\approx0.3$ keV 
(cooler than typical accretion disks of Galactic black holes).
The characteristic inner-disk radius implied 
by the thermal component is $\approx1000$ km: 
this corresponds to a Schwarzschild black hole mass $\approx100\, M_{\odot}$ 
if we are directly seeing the disk all the way to the innermost stable circular orbit, 
or $< 100\, M_{\odot}$ if the inner disk is hidden by a Comptonizing region. 
The latter scenario is more likely, given the dominant power-law component.
In fact, in some \chandra\ observations the thermal component disappears altogether, 
which suggests that a larger fraction of the disk emission is Comptonized at such times.

A strict application of the Eddington limit 
requires a black hole mass $\ga40\,M_{\odot}$; 
however, luminosities up to $\sim3$ times Eddington 
(implying correspondingly lower black hole masses) 
can be produced in standard accretion scenarios.
For example, analytical solutions of standard accretion models 
show that the true emitted luminosity $\sim(1 + \log\dot{M})$ for $\dot{M} > 1$
\citep{plfba2007}.
Recent radiation-magnetohydrodynamic simulations confirm 
that the isotropic luminosity can reach $\sim L_{edd}$ 
during super-critical accretion \citep{om2011}.
In addition, mild geometrical beaming can further increase the apparent luminosity
by a factor of two for a standard disk seen face-on
and by a factor of 10 for supercritical accretion flows \citep{om2011}.
Observationally, several neutron star X-ray binaries 
have reached luminosities of a few times Eddington 
in their flaring state \citep{bgjc2010,barnardea2003,homanea2007}.
Thus, calculated black hole masses may be overestimated by a factor of $\la3$.

It is still an unsolved theoretical problem whether black holes
can form from standard stellar evolution at high metallicity 
\citep{belczynski10, heger03}, 
as metal lines make mass loss more efficient, 
decreasing the pre-collapse stellar mass drastically.
As pointed out in \citet{heger03}, there are very large uncertainties
on exactly what mass/metallicities can actually produce a black hole.
It has been argued \citep{pakull02, mapelli09, zampieri09}
that ULX formation either requires a low-metallicity environment,
or is enhanced by a low-metallicity environment \citep{prestwich11}.
This source demonstrates that low metallicity need not always be the case, 
since the environment in the inner disk of M83 where the ULX is located 
can hardly be characterized as a low-metallicity one.
Accurate measurements of abundances in regions with relatively high metallicity 
are notoriously difficult, 
but recent estimates based on deep spectra of \ion{H}{2} regions 
within the D$_{25}$ radius give oxygen abundances
for a radius of $1\farcm0$ from the nucleus to be
12+log(O/H)=8.73$\pm0.01$ \citep{bresolin09},
compared to the solar value of 8.69 \citep{ags2005}
and the local ISM value of 8.69 \citep{snowwitt1996}.
A different method applied to the \ion{H}{2} 
regions of M83 produces for the same radius
12+log(O/H)=8.59$\pm0.01$ \citep{pilyugin2006}
compared to their local ISM reference value of 8.50.
While there is some disagreement about the absolute oxygen abundance,
it is clear that the relative oxygen abundance for this region
is slightly higher than solar.
In such an environment, 
even an older population is unlikely to have a low metallicity, 
unless the ULX is from a disrupted dwarf galaxy.
This object suggests that black holes with masses $\ga40\,M_{\odot}$
can be found in environments where the local abundance is solar to somewhat super-solar,
whereas models suggest that at these abundances 
only black holes with $\la15\,M_{\odot}$ are produced \citep{belczynski10}.
Reducing abundances to LMC values can produce black holes $\la30\,M_{\odot}$,
so finding a $\ga40\,M_{\odot}$ black hole in a region with solar abundances,
even accounting for enrichment since its stellar formation,
is rather unusual and difficult to understand.

{\subsection{The Nature of the Optical Counterpart} 

The optical colors of the system in outburst
are not consistent with a simple blackbody-like spectrum;
the flux density, as shown in Fig.\ \ref{fig:opt_xray}, 
decreases from the near-UV to the visible band, 
but increases again towards the I band. 
This behavior suggests the presence of at least two components:
one peaking in the UV (implying a characteristic blackbody temperature
$\ga 20,000$ K and a characteristic effective radius $\la 20 R_{\odot}$)
and one in the IR (implying a characteristic blackbody temperature
$\la 4,000$ K and a characteristic effective radius $\ga 100 R_{\odot}$).
There are (at least) two ways to interpret this situation. 
One possibility is that the IR emission is dominated by the large outer disk, 
and the UV excess is the hot spot or the irradiated surface of the donor star. 
An alternative is that the UV peak is 
the Rayleigh-Jeans tail of the emission from an irradiated disk 
and the I-band excess is the Wien tail 
of the emission from a cool stellar component.
In the latter case the disk must be truncated 
or at least shaded from irradiation beyond $R\approx20R_{\odot}$. 
If the disk is shaded, there will be no significant emission 
from beyond $R\approx20R_{\odot}$ as the temperature of a non-irradiated disk
would be 2000-3000 K \citep{shakura73,fkr2002}
and would drop rapidly with distance.
At those temperatures, the disk gas is entirely neutral, and opacity
drops to a minimum \citep{ferguson05}. It is still debated whether the outer disk is optically thin or thick to the continuum in that regime \cite[{e.g.}][]{cannizzo84, hynes05, idan10}; however, in either case, its emission should be negligible compared to that from the irradiated part of the disk and the donor star.
The pre-outburst optical spectral energy distribution contains only a red component
which we suggest is either the donor star 
or an unrelated red star along the line of sight.
When we add to that an irradiated disk model, 
with disk parameters constrained by the {\it Chandra} observations of March 2011,
we recover the optical brightnesses of the 2011 {\it HST} observations.
Thus we argue that UV/optical emission is due primarily to the irradiated disk,
with some stellar contribution in the near-IR.

We used the XSPEC model {\it diskir} \citep{gierlinski2009,gierlinski2008}
to model the irradiated disk emission for the 23 March 2011 \chandra\ data.
This observation was chosen because it was the longest,
had the best constrained accretion disk parameters,
and was made close to the Gemini observations,
which we can then tie to the \hst\ observations.
We verified that the best-fitting disk and power-law parameters 
obtained with {\it diskir} for the X-ray data
are consistent with the corresponding parameters obtained with our previous fits.
The advantage of the {\it diskir} model over the simpler, 
more traditional {\it diskbb} plus {\it power-law} model 
is that it allows a more natural extrapolation into the optical band, 
as the power-law component does not extend beyond the peak of the disk emission.
We found that the UV/optical colors of the 2011 \hst\ observations
are best-fitted with a disk truncation radius $r_{\rm out} \approx
11,200 ~ r_{\rm in} \approx 1.4 \times 10^{12}$ cm $\approx 20 R_{\odot}$,
and a fraction of X-ray photons intercepted and re-radiated by the disk
to be $\approx 5 \times 10^{-3}$, 
a very plausible value for irradiated disks in binary systems. 
Reprocessing fractions between $\approx 10^{-3}$ and 
$\approx 10^{-2}$ are suggested both by theoretical modelling 
\citep[e.g.,][]{vrtilek90,dejong96,king97,dubus99} and by observations 
of accretion disks in Galactic black holes 
\citep[e.g.,][]{hynes02,gierlinski2009}.

The red excesses in both the pre- and post-outburst
spectral energy distributions (SEDs)
can be represented as a blackbody with $T \approx 3500$ K
and $\log L/L_{\odot} \approx 3.5$, implying a radius $\approx 150 R_{\odot}$.
This set of parameters is consistent with an AGB star of initial mass
$\approx 2.5$--$4 M_{\odot}$ and an age $\approx 200$--$800$ Myr
\citep[Padova stellar tracks:][]{salasnich00,bertelli08,bertelli09,marigo08,girardi10}.
We do not have enough evidence to determine whether the faint red component
comes from the ULX donor star, or from an unrelated evolved star 
that is projected in the sky within $0.1\arcsec$ of the X-ray source,
or even from a positive fluctuation in the background stellar density.
There are several other faint red stars in the field (Figure~\ref{fig:wfc3}).
If some of the red emission seen from the optical counterpart 
when the ULX is in outburst stems from an unrelated star, 
then the true donor star of the ULX would be even fainter and less massive
than we have calculated, but our interpretation of the system remains valid. 
Even if the red excess is due to a single stellar companion,
the parameters given above are descriptive rather than prescriptive;
binary evolution is likely to have produced a companion whose structure
is somewhat different than an isolated AGB star.

The upper mass/luminosity limits that we place on the optical counterpart
are not inconsistent with population studies of the surrounding region.
Hwihyun Kim (private comunication) 
provided photometry, reddening, and age estimates
for the bright stars in a region around the ULX, 
based on an analysis of the color-color and color-magnitude diagrams.
However, none of those stars is close to the ULX ($r>60$ pc);
the population surrounding the ULX is dominated by an older 
and fainter population than is accessible 
by color-color and color-magnitude diagrams, even with \hst .

If we had observed the optical counterpart {\em only} 
in the X-ray luminous state, 
we would have classified it as a main-sequence early-type 
B star or a blue supergiant, 
in agreement with most other ULXs for which a unique optical counterpart 
has been proposed \citep{tao11}. 
However, in at least one case \citep{feng08,roberts08},
it was already suspected that the optical brightness
is strongly affected by X-ray irradiation and reprocessing on the stellar
surface and/or the outer accretion disk. 
Here, for the first time we can prove that this is the case 
because this source is a transient 
and we have two sets of observations 
in which the optical brightness is strongly different.
It is plausible that some other ULXs with apparently massive donor stars 
in fact have a low-mass donor and a bright accretion disk.

{\subsection{The Geometry of the System} 

The high mass-accretion rate required by the X-ray luminosity, 
coupled with the low soft-X-ray absorption observed in the X-ray data, 
imply mass transfer via Roche lobe overflow 
(as should be the case for most ULXs), 
and hence the formation of a large accretion disk around the black hole.
We have argued that most of the optical emission in outburst is likely to come 
from the irradiated disk. The required fraction of intercepted X-ray photons 
is comparable to the typical values inferred for Galactic LMXBs, for which 
the emission is thought to be quasi-isotropic.
We would not expect the same level of reprocessing if most of the X-ray luminosity 
was beamed along the axis perpendicular to the disk plane. 
Thus the presence of strong optical reprocessing
is further evidence that a ULX is {\em not} strongly beamed.

Quantitative estimates of the outer disk size and temperature 
based on broad-band fits to the X-ray/optical spectral energy distribution 
are somewhat model-dependent; 
the disk-blackbody component of the X-ray emission varies from epoch to epoch.
However, we can obtain useful order-of-magnitude estimates of the system size 
by applying the empirical relation between optical brightness 
of an irradiated disk and binary period found by \citet{van-paradijs94}: 
$$M_V = 1.57 - 1.51 \log P_h - 1.14 \log(L_{\rm X}/L_{\rm Edd}),$$
where $P_h$ is the binary period in hours. 
In outburst, the absolute visual brightness 
of the ULX counterpart in the standard (Vegamag) system
is $M_V \approx -4.9$ mag (Table~\ref{tab:opti}, 
after correcting for line-of-sight extinction). 
If the X-ray luminosity $L_{\rm X} \approx (1$--$3) \times L_{\rm Edd}$, 
as most classes of ULX models suggest \citep[e.g.][]{feng11},
this corresponds to an orbital period $\approx 360$ to $830$ days.
Since the donor star is filling its Roche lobe,
$P_{\rm h}\,\rho^{1/2} \approx 10.5$ for mass ratios
$0.01 \la q \la 1$ \citep{eggleton83}, 
where $\rho$ is the average stellar density in g cm$^{-3}$. 
Thus, we obtain $3 \times 10^{-7} \la (\rho/{\rm g~cm}^{-3}) \la 1.5 \times 10^{-6}$.
These are typical densities of red giants or AGB stars
\citep{salasnich00, bertelli08, bertelli09}, 
consistent with our interpretation of the system. 
As a comparison, our independent estimate of a mass 
$\approx 3 M_{\odot}$ and radius $\approx 150 R_{\odot}$
for the red star apparently associated with the ULX implies a mean density
$\rho \approx 10^{-6}$ g cm$^{-3}$, 
in agreement with the expectations if that is the true donor star.

The size of the accretion disk remains uncertain. 
Using our lowest estimate for M$_{BH}\sim40$ M$_{\odot}$
and our upper estimate for the M$_{star}\sim4$ M$_{\odot}$,
the radius of the primary's Roche lobe is at least a few hundred $R_{\odot}$.
The maximum size of an accretion disk is $\approx 80\%$ of the size
of the primary's Roche lobe \citep{paczynski77, whitehurst88},
which corresponds to a characteristic size $\approx 10^{13}$ cm for this ULX. 
On the other hand, we have argued that the optical colors
and luminosity in outburst suggest an irradiated disk size $\approx 10^{12}$ cm. 
One possibility is that the disk is truncated at that radius, 
in which case the whole disk would be kept ionized by the X-ray illumination. 
Alternatively, the outer part of the disk may be shaded from irradiation, 
and therefore much colder, neutral, 
and not a significant contributor to the optical/IR emission;
the temperature of a non-irradiated standard disk \citep{shakura73}
at $R \approx 10^{13}$ cm would be only $\approx 600$ K, 
for the disk parameters fitted to the X-ray emission.
We do not have enough information to choose between the two scenarios.
Whether the outer disk is mostly ionized or neutral has implications
for the transient behavior.

{\subsection{Transient Behavior} 

Most Galactic black holes with a low-mass donor star are transient, 
while most of those with an OB donor are persistent X-ray emitters 
(albeit with flux and spectral variability). 
We suggest that the same may be true for ULXs. 
A few other examples of ULX transients are two sources 
in NGC\,1365 \citep{soria09}, one in M\,31 \citep{kaur11}
two in M\,101 \citep{kuntz05}, one in NGC\,3628 \citep{strickland01},
and one in M\,82 \citep{feng07}.
The transient behavior in accreting stellar-mass black holes 
has been attributed to thermal/viscous disk instabilities 
\citep{mineshige89, osaki96, lasota01}
and/or to mass transfer instabilities
\citep{hameury86, hameury87, hameury88, viallet08}.

Mass-transfer instabilities operate when the evolved donor star is nearly
but not entirely filling its Roche lobe, 
providing only a low accretion rate 
and thus producing a low-luminosity ``quiescent'' state. 
The resulting X-ray photons penetrate a few Thompson optical depths 
below the photosphere of the donor star, 
causing the convective envelope to expand slowly
until the star makes full contact with its Roche lobe.
This contact dramatically enhances both the mass transfer rate through the L1 point
and the X-ray luminosity of the system, which is generally seen as an ``outburst'' state. 
However, as a larger accretion disk is formed, 
it eventually shades the L1 region from X-ray irradiation, 
decreasing or stopping mass transfer.
The parameter space for the mass 
transfer instability in ULXs is largely unexplored.
A necessary condition is that the irradiating X-ray flux at the surface of the donor 
star be stronger than its intrinsic flux. 
This is not the case for ULXs with massive donors, 
but is true for this source and probably for other ULXs with low-mass donors.

In the disk instability model, the disk follows a limit cycle
between a hot (higher viscosity) and a cold (lower viscosity) state.
The difference between the two states is due to the sudden jump in opacity
when the temperature reaches $\approx 6,500$ K and hydrogen becomes mostly ionized
\citep{cannizzo88}.
The instability can be suppressed if the whole disk is kept ionized
by X-ray irradiation \citep{burderi98, king98, janiuk11}. 
For the observed luminosity of the M83 ULX at the peak of the outburst,
the outer disk can be kept ionized by irradiation up to a radius
$R_{\rm irr} \approx 10^{13} (f_{\rm irr}/10^{-2})^{1/2}$ cm, 
where $f_{\rm irr}$ is the fraction of X-ray flux reprocessed by the disk.
From the inferred size of the binary system,
the accretion disk can extend beyond this instability boundary
if the outer radius reaches the tidal truncation radius
\citep{paczynski77, whitehurst88}.  
On the other hand, the observed optical colors suggest that
the outer disk of the ULX has a temperature $\ga 20,000$ K 
and hence is truncated well inside the tidal radius.
Thus, we cannot tell whether the disk instability operates on the ULX in M83
and will be responsible for ultimately bringing the outburst to a close.

If the outburst does end soon, 
and the X-ray luminosity declines to below $10^{39}$ erg s$^{-1}$, 
we will have a great opportunity to monitor its spectral behavior 
in the luminosity range typical of Galactic black holes.
Determining whether or not a ULX behaves like an ordinary stellar-mass 
black hole (e.g., similar state transitions and evolution 
in the hardness-intensity diagram) when its luminosity drops will tell us 
whether it is powered by an intrinsically different type of black hole, 
or by an ordinary stellar-mass black hole at an extremely high accretion rate, 
not usually reached by Galactic black holes.

\section{Summary}

We have discovered a new ULX in M83 using \chandra, 
and have characterized the X-ray properties 
of the source in a series of \chandra\ and \swift\ 
observations extending through December 2011.
We have also detected the optical counterpart to the source 
using the Gemini South Telescope and \hst. 
The ULX is located in an inter-arm region 
and well away from sites of active star formation. 
Its observed properties and our interpretation of them can be summarized as follows:

\begin{enumerate}
\item
At its discovery in December 2010, 
the luminosity was $L_X (0.3 - 10\,{\rm keV}) \approx 4 \times 10^{39} \LUM$.  
The source has remained bright; 
its luminosity has varied by a factor of two,  
and it has only recently dropped to $\approx 2 \times 10^{39} \LUM$.
There is no previous evidence for the existence of this X-ray source 
in observations extending back to 1979; 
the X-ray flux has increased by at least a factor of 3000. 

\item
Although there is significant variation between observations,
the X-ray light-curves of individual observations 
show no signs of short-term variability,
nor are there signs of orbital modulations or eclipses.

\item
The X-ray spectra can be well fitted by a disk blackbody plus power-law model, 
or by an absorbed Comptonized spectrum, typical of most ULXs.  
The strength of the disk blackbody component varies on timescales of days. 
We  attribute this to the fraction of inner disk photons upscattered 
in a variable Comptonizing region, rather than to a change in the disk.  

\item
In those X-ray spectra where a disk is evident, 
the disk luminosity and temperature suggest 
an inner accretion disk radius of about 1000 km, 
corresponding to a Schwarzschild black hole with 
mass $M_{BH} \approx 100\, M_\sun$.
The black hole mass could be less 
if the inner disk is hidden by a Comptonizing region.
In order to strictly obey the Eddington limit, 
the black hole must have a mass $M_{BH} \gtrsim 40\, M_\sun$.
However, if we allow the possibility of accretion luminosities 
up to three times the Eddington limit at super-critical accretion rates, 
the X-ray data are still consistent with an ordinary stellar-mass black hole.

\item
A blue optical counterpart with $M_V \approx -4.85$
has appeared since August 2009, presumably at the same time as the ULX.
The only stars near this site 
prior to the appearance of the ULX are faint and red,
so must belong to an older population.
The donor star is not an OB star,
and is likely to be a red giant or AGB star with $M \lesssim 4\, M_\sun$
and age $\ga 500$ Myr.
We note, however, that had the system been observed only in its luminous state,
the donor could easily have been mistakenly interpreted as an OB star,
consistent with most other ULXs with uniquely identified optical counterparts.
Some of these other ULXs may well have low-mass donor stars as well.

\item
During the X-ray outburst, the spectral energy distribution
of the optical counterpart is dominated by a blue component,
with $M_V \approx -4.85$ mag, which we interpret as
the Rayleigh-Jeans tail of the emission from the outer disk,
heated by the X-ray photons.
In addition, there is a faint red component
that may arise from the surface of the donor star,
although we cannot exclude the possibility that it stems
from other unrelated stars in the vicinity.

\item
The M83 ULX system provides clear evidence that not all ULXs 
involve an OB donor and a young stellar population, 
confirming the suggestion from statistical studies that there are two classes of ULXs.

\item
The existence of a ULX in the inner disk of M83 suggests that it is possible 
to produce black holes also in systems with at least solar metallicity.

\end{enumerate}

The ULX in M83 has provided significant insight into the diversity of the ULX population,
and provides us with an unequivocal example of a ULX with a low-mass donor.
Continued monitoring of this source will bring greater depth of understanding
to this source and this class of sources.
Determining the length of the current outburst
will constrain the extent to which sources like this
contribute to the persistent ULX population,
and the length of the decline may constrain the size of the accretion disk.
It is still actively debated whether
ULXs are a state reached by ordinary stellar-mass black holes at extremely high
(super-Eddington) accretion rates,
or are powered by a different (more massive) type of accreting black hole.
Almost all Galactic black hole transients pass through a high/soft state
(dominated by thermal disk emission)
after the peak of their outburst and before returning to quiescence;
they remain in that state for several weeks.
If this ULX does the same, the luminosity at which the state changes
will help us understand and quantify the relation 
between stellar-mass black holes and ULXs.
If, instead, this ULX continues to behave like a ULX even
when its luminosity goes below $\approx10^{39}$ erg s$^{-1}$,
then it is likely that there are intrinsic physical differences between
the BHs in ULXs and the ordinary stellar-mass black holes.
Clearly, this is an interesting object that will continue
to illuminate our ignorance about ULXs.

\acknowledgements

We would like to thank Neil Gehrels and the \swift\ team
for approving and supporting our three series of \swift\ DDT observations,
and for their flexibility and cooperation in scheduling observations 
to coincide with those from \chandra\ and \hst.
We would like to thank Matt Mountain, Director of the  Space Telescope Science Institute 
for granting us Director's Discretionary time for the second-epoch WFC3 observation.   
We thank Hwihyun Kim (Arizona State University)
for providing photometry of the HST data in the field surrounding the ULX position.
We would also like to thank the referee for a useful referee's report.

Support for this work was provided by the National Aeronautics and Space Administration 
through \chandra\ Grant Number G01-12115, issued by the \chandra\ X-ray Observatory Center, 
which is operated by the Smithsonian Astrophysical Observatory 
for and on behalf of NASA under contract NAS8-03060\@.   
PFW and WPB are grateful for both observing and travel support 
for the Gemini observations from the Gemini office at NOAO.  
PFW also  acknowledges financial support from the National Science Foundation 
through grant AST-0908566, 
and the hospitality of the Research School of Astronomy and Astrophysics, 
Australian National University, during a portion of the work presented here.

We acknowledge the use of public data from the Swift data archive.
This research has made use of data obtained from the 
High Energy Astrophysics Science Archive Research Center (HEASARC), 
provided by NASA's Goddard Space Flight Center.

The Gemini Observatory is operated by AURA under a cooperative agreement 
with the NSF on behalf of the Gemini partnership: the National Science Foundation (United 
States), the Science and Technology Facilities Council (United Kingdom), the 
National Research Council (Canada), CONICYT (Chile), the Australian Research Council (Australia), 
Ministerio da Ciencia, Tecnologia e Inovation (Brazil) 
and Ministerio de Ciencia, Tecnologia e Innovacion Productiva (Argentina).


\begin{thebibliography}{}

\bibitem[\protect\citeauthoryear{{Arnaud}}{{Arnaud}}{1996}]{arnaud96}
{Arnaud}, K.~A. 1996, in Astronomical Society of the Pacific Conference Series,
  Vol. 101, Astronomical Data Analysis Software and Systems V, ed.
  {G.~H.~Jacoby \& J.~Barnes}, 17

\bibitem[\protect\citeauthoryear{{Asplund}, {Grevesse}, \& {Sauval}}{{Asplund}
  et~al.}{2005}]{ags2005}
{Asplund}, M., {Grevesse}, N.,  \& {Sauval}, A.~J. 2005, in Astronomical
  Society of the Pacific Conference Series, Vol. 336, Cosmic Abundances as
  Records of Stellar Evolution and Nucleosynthesis, ed. {T.~G.~Barnes III \&
  F.~N.~Bash}, 25

\bibitem[\protect\citeauthoryear{{Ba{\l}uci{\'n}ska-Church}
  et~al.}{{Ba{\l}uci{\'n}ska-Church} et~al.}{2010}]{bgjc2010}
{Ba{\l}uci{\'n}ska-Church}, M., {Gibiec}, A., {Jackson}, N.~K.,  \& {Church},
  M.~J. 2010, \aap, 512, A9

\bibitem[\protect\citeauthoryear{{Barnard}, {Church}, \&
  {Ba{\l}uci{\'n}ska-Church}}{{Barnard} et~al.}{2003}]{barnardea2003}
{Barnard}, R., {Church}, M.~J.,  \& {Ba{\l}uci{\'n}ska-Church}, M. 2003, \aap,
  405, 237

\bibitem[\protect\citeauthoryear{{Bauer} et~al.}{{Bauer}
  et~al.}{2004}]{bauer04}
{Bauer}, F.~E., {Alexander}, D.~M., {Brandt}, W.~N., {Schneider}, D.~P.,
  {Treister}, E., {Hornschemeier}, A.~E.,  \& {Garmire}, G.~P. 2004, \aj, 128,
  2048

\bibitem[\protect\citeauthoryear{{Begelman}}{{Begelman}}{2002}]{begelman02}
{Begelman}, M.~C. 2002, \apjl, 568, L97

\bibitem[\protect\citeauthoryear{{Begelman}, {King}, \& {Pringle}}{{Begelman}
  et~al.}{2006}]{begelman06}
{Begelman}, M.~C., {King}, A.~R.,  \& {Pringle}, J.~E. 2006, \mnras, 370, 399

\bibitem[\protect\citeauthoryear{{Belczynski} et~al.}{{Belczynski}
  et~al.}{2010}]{belczynski10}
{Belczynski}, K., {Bulik}, T., {Fryer}, C.~L., {Ruiter}, A., {Valsecchi}, F.,
  {Vink}, J.~S.,  \& {Hurley}, J.~R. 2010, \apj, 714, 1217

\bibitem[\protect\citeauthoryear{{Berghea} et~al.}{{Berghea}
  et~al.}{2008}]{berghea08}
{Berghea}, C.~T., {Weaver}, K.~A., {Colbert}, E.~J.~M.,  \& {Roberts}, T.~P.
  2008, \apj, 687, 471

\bibitem[\protect\citeauthoryear{{Bertelli} et~al.}{{Bertelli}
  et~al.}{2008}]{bertelli08}
{Bertelli}, G., {Girardi}, L., {Marigo}, P.,  \& {Nasi}, E. 2008, \aap, 484,
  815

\bibitem[\protect\citeauthoryear{{Bertelli} et~al.}{{Bertelli}
  et~al.}{2009}]{bertelli09}
{Bertelli}, G., {Nasi}, E., {Girardi}, L.,  \& {Marigo}, P. 2009, \aap, 508,
  355

\bibitem[\protect\citeauthoryear{{Blackburn}}{{Blackburn}}{1995}]{blackburn95}
{Blackburn}, J.~K. 1995, in Astronomical Society of the Pacific Conference
  Series, Vol.~77, Astronomical Data Analysis Software and Systems IV, ed.
  {R.~A.~Shaw, H.~E.~Payne, \& J.~J.~E.~Hayes}, 367

\bibitem[\protect\citeauthoryear{{Bresolin} et~al.}{{Bresolin}
  et~al.}{2009}]{bresolin09}
{Bresolin}, F., {Ryan-Weber}, E., {Kennicutt}, R.~C.,  \& {Goddard}, Q. 2009,
  \apj, 695, 580

\bibitem[\protect\citeauthoryear{{Burderi}, {King}, \&
  {Szuszkiewicz}}{{Burderi} et~al.}{1998}]{burderi98}
{Burderi}, L., {King}, A.~R.,  \& {Szuszkiewicz}, E. 1998, \apj, 509, 85

\bibitem[\protect\citeauthoryear{{Caballero-Garc{\'{\i}}a} \&
  {Fabian}}{{Caballero-Garc{\'{\i}}a} \& {Fabian}}{2010}]{caballero10}
{Caballero-Garc{\'{\i}}a}, M.~D.,  \& {Fabian}, A.~C. 2010, \mnras, 402, 2559

\bibitem[\protect\citeauthoryear{{Cannizzo}, {Shafter}, \&
  {Wheeler}}{{Cannizzo} et~al.}{1988}]{cannizzo88}
{Cannizzo}, J.~K., {Shafter}, A.~W.,  \& {Wheeler}, J.~C. 1988, \apj, 333, 227

\bibitem[Cannizzo \& Wheeler(1984)]{cannizzo84} Cannizzo, J.~K., \& Wheeler, J.~C.\ 1984, \apjs, 55, 367 

\bibitem[\protect\citeauthoryear{{Cappelluti} et~al.}{{Cappelluti}
  et~al.}{2009}]{cappelluti09}
{Cappelluti}, N., et~al. 2009, \aap, 497, 635

\bibitem[\protect\citeauthoryear{{Cardelli}, {Clayton}, \& {Mathis}}{{Cardelli}
  et~al.}{1989}]{cardelli89}
{Cardelli}, J.~A., {Clayton}, G.~C.,  \& {Mathis}, J.~S. 1989, \apj, 345, 245

\bibitem[\protect\citeauthoryear{{Castro-Tirado}}{{Castro-Tirado}}{2011}]{castro-tirado11}
{Castro-Tirado}, A.~J. 2011, in IAU Symposium, Vol. 275, IAU Symposium, ed.
  {G.~E.~Romero, R.~A.~Sunyaev, \& T.~Belloni}, 270

\bibitem[\protect\citeauthoryear{{Colbert} et~al.}{{Colbert}
  et~al.}{2004}]{colbert04}
{Colbert}, E.~J.~M., {Heckman}, T.~M., {Ptak}, A.~F., {Strickland}, D.~K.,  \&
  {Weaver}, K.~A. 2004, \apj, 602, 231

\bibitem[\protect\citeauthoryear{{Colbert} \& {Mushotzky}}{{Colbert} \&
  {Mushotzky}}{1999}]{colbert99}
{Colbert}, E.~J.~M.,  \& {Mushotzky}, R.~F. 1999, \apj, 519, 89

\bibitem[\protect\citeauthoryear{{Copperwheat} et~al.}{{Copperwheat}
  et~al.}{2007}]{copperwheat07}
{Copperwheat}, C., {Cropper}, M., {Soria}, R.,  \& {Wu}, K. 2007, \mnras, 376,
  1407

\bibitem[\protect\citeauthoryear{{Davis}}{{Davis}}{2001}]{davis2001}
{Davis}, J.~E. 2001, \apj, 562, 575

\bibitem[\protect\citeauthoryear{{de Jong}, {van Paradijs}, \&
  {Augusteijn}}{{de Jong} et~al.}{1996}]{dejong96}
{de Jong}, J.~A., {van Paradijs}, J.,  \& {Augusteijn}, T. 1996, \aap, 314, 484

\bibitem[\protect\citeauthoryear{{Dewangan}, {Titarchuk}, \&
  {Griffiths}}{{Dewangan} et~al.}{2006}]{dewangan06}
{Dewangan}, G.~C., {Titarchuk}, L.,  \& {Griffiths}, R.~E. 2006, \apjl, 637,
  L21

\bibitem[\protect\citeauthoryear{{Dopita} et~al.}{{Dopita}
  et~al.}{2010}]{dopm83}
{Dopita}, M.~A., et~al. 2010, \apj, 710, 964

\bibitem[\protect\citeauthoryear{{Dubus} et~al.}{{Dubus}
  et~al.}{1999}]{dubus99}
{Dubus}, G., {Lasota}, J.-P., {Hameury}, J.-M.,  \& {Charles}, P. 1999, \mnras,
  303, 139

\bibitem[\protect\citeauthoryear{{Eggleton}}{{Eggleton}}{1983}]{eggleton83}
{Eggleton}, P.~P. 1983, \apj, 268, 368

\bibitem[\protect\citeauthoryear{{Fender}, {Belloni}, \& {Gallo}}{{Fender}
  et~al.}{2004}]{fender04}
{Fender}, R.~P., {Belloni}, T.~M.,  \& {Gallo}, E. 2004, \mnras, 355, 1105

\bibitem[\protect\citeauthoryear{{Feng} \& {Kaaret}}{{Feng} \&
  {Kaaret}}{2007}]{feng07}
{Feng}, H.,  \& {Kaaret}, P. 2007, \apj, 668, 941

\bibitem[\protect\citeauthoryear{{Feng} \& {Kaaret}}{{Feng} \&
  {Kaaret}}{2008}]{feng08}
{Feng}, H.,  \& {Kaaret}, P. 2008, \apj, 675, 1067

\bibitem[\protect\citeauthoryear{{Feng} \& {Soria}}{{Feng} \&
  {Soria}}{2011}]{feng11}
{Feng}, H.,  \& {Soria}, R. 2011, ArXiv e-prints

\bibitem[Ferguson et al.(2005)]{ferguson05} Ferguson, J.~W., Alexander, D.~R., Allard, F., Barman, T., Bodnarik, J.~G., 
Hauschildt, P.~H., Heffner-Wong, A., \& Tamanai, A.\ 2005, \apj, 623, 585 

\bibitem[\protect\citeauthoryear{{Frank, J., King, A., \& Raine,
  D.~J.}}{{Frank, King \& Raine}}{2002}]{fkr2002}
{Frank, J., King, A., \& Raine, D.~J.}, ed. 2002, {Accretion Power in
  Astrophysics: Third Edition}

\bibitem[\protect\citeauthoryear{{Fruscione} et~al.}{{Fruscione}
  et~al.}{2006}]{fruscione06}
{Fruscione}, A., et~al. 2006, in Society of Photo-Optical Instrumentation
  Engineers (SPIE) Conference Series, Vol. 6270, Society of Photo-Optical
  Instrumentation Engineers (SPIE) Conference Series

\bibitem[\protect\citeauthoryear{{Gierli{\'n}ski}, {Done}, \&
  {Page}}{{Gierli{\'n}ski} et~al.}{2008}]{gierlinski2008}
{Gierli{\'n}ski}, M., {Done}, C.,  \& {Page}, K. 2008, \mnras, 388, 753

\bibitem[\protect\citeauthoryear{{Gierli{\'n}ski}, {Done}, \&
  {Page}}{{Gierli{\'n}ski} et~al.}{2009}]{gierlinski2009}
{Gierli{\'n}ski}, M., {Done}, C.,  \& {Page}, K. 2009, \mnras, 392, 1106

\bibitem[\protect\citeauthoryear{{Girardi} et~al.}{{Girardi}
  et~al.}{2010}]{girardi10}
{Girardi}, L., et~al. 2010, \apj, 724, 1030

\bibitem[\protect\citeauthoryear{{Gladstone}, {Roberts}, \& {Done}}{{Gladstone}
  et~al.}{2009}]{gladstone09}
{Gladstone}, J.~C., {Roberts}, T.~P.,  \& {Done}, C. 2009, \mnras, 397, 1836

\bibitem[\protect\citeauthoryear{{Goad} et~al.}{{Goad} et~al.}{2006}]{goad06}
{Goad}, M.~R., {Roberts}, T.~P., {Reeves}, J.~N.,  \& {Uttley}, P. 2006,
  \mnras, 365, 191

\bibitem[\protect\citeauthoryear{{Gris{\'e}} et~al.}{{Gris{\'e}}
  et~al.}{2008}]{grise08}
{Gris{\'e}}, F., {Pakull}, M.~W., {Soria}, R., {Motch}, C., {Smith}, I.~A.,
  {Ryder}, S.~D.,  \& {B{\"o}ttcher}, M. 2008, \aap, 486, 151

\bibitem[\protect\citeauthoryear{{Hameury}, {King}, \& {Lasota}}{{Hameury}
  et~al.}{1986}]{hameury86}
{Hameury}, J.~M., {King}, A.~R.,  \& {Lasota}, J.~P. 1986, \aap, 162, 71

\bibitem[\protect\citeauthoryear{{Hameury}, {King}, \& {Lasota}}{{Hameury}
  et~al.}{1987}]{hameury87}
{Hameury}, J.~M., {King}, A.~R.,  \& {Lasota}, J.~P. 1987, \aap, 171, 140

\bibitem[\protect\citeauthoryear{{Hameury}, {King}, \& {Lasota}}{{Hameury}
  et~al.}{1988}]{hameury88}
{Hameury}, J.~M., {King}, A.~R.,  \& {Lasota}, J.~P. 1988, Advances in Space
  Research, 8, 489

\bibitem[\protect\citeauthoryear{{Hamuy} et~al.}{{Hamuy}
  et~al.}{1992}]{hamuy92}
{Hamuy}, M., {Walker}, A.~R., {Suntzeff}, N.~B., {Gigoux}, P., {Heathcote},
  S.~R.,  \& {Phillips}, M.~M. 1992, \pasp, 104, 533

\bibitem[\protect\citeauthoryear{{Heger} et~al.}{{Heger}
  et~al.}{2003}]{heger03}
{Heger}, A., {Fryer}, C.~L., {Woosley}, S.~E., {Langer}, N.,  \& {Hartmann},
  D.~H. 2003, \apj, 591, 288

\bibitem[\protect\citeauthoryear{{Heil}, {Vaughan}, \& {Roberts}}{{Heil}
  et~al.}{2009}]{heil09}
{Heil}, L.~M., {Vaughan}, S.,  \& {Roberts}, T.~P. 2009, \mnras, 397, 1061

\bibitem[\protect\citeauthoryear{{Homan} et~al.}{{Homan}
  et~al.}{2007}]{homanea2007}
{Homan}, J., et~al. 2007, \apj, 656, 420

\bibitem[\protect\citeauthoryear{{Hornschemeier} et~al.}{{Hornschemeier}
  et~al.}{2001}]{hornschemeier01}
{Hornschemeier}, A.~E., et~al. 2001, \apj, 554, 742

\bibitem[\protect\citeauthoryear{{Hynes} et~al.}{{Hynes}
  et~al.}{2002}]{hynes02}
{Hynes}, R.~I., {Haswell}, C.~A., {Chaty}, S., {Shrader}, C.~R.,  \& {Cui}, W.
  2002, \mnras, 331, 169
  
\bibitem[Hynes et al.(2005)]{hynes05} Hynes, R.~I., Robinson, E.~L., \& Bitner, M.\ 2005, \apj, 630, 405 

\bibitem[Idan et al.(2010)]{idan10} Idan, I., Lasota, J.-P., Hameury, J.-M., \& Shaviv, G.\ 2010, \aap, 519, A117 

\bibitem[\protect\citeauthoryear{{Immler} et~al.}{{Immler}
  et~al.}{1999}]{immler99}
{Immler}, S., {Vogler}, A., {Ehle}, M.,  \& {Pietsch}, W. 1999, \aap, 352, 415

\bibitem[\protect\citeauthoryear{{Irwin}, {Athey}, \& {Bregman}}{{Irwin}
  et~al.}{2003}]{irwin03}
{Irwin}, J.~A., {Athey}, A.~E.,  \& {Bregman}, J.~N. 2003, \apj, 587, 356

\bibitem[\protect\citeauthoryear{{Irwin}, {Bregman}, \& {Athey}}{{Irwin}
  et~al.}{2004}]{irwin04}
{Irwin}, J.~A., {Bregman}, J.~N.,  \& {Athey}, A.~E. 2004, \apjl, 601, L143

\bibitem[\protect\citeauthoryear{{Janiuk} \& {Czerny}}{{Janiuk} \&
  {Czerny}}{2011}]{janiuk11}
{Janiuk}, A.,  \& {Czerny}, B. 2011, \mnras, 414, 2186

\bibitem[\protect\citeauthoryear{{Jonker} \& {Nelemans}}{{Jonker} \&
  {Nelemans}}{2004}]{jonker04}
{Jonker}, P.~G.,  \& {Nelemans}, G. 2004, \mnras, 354, 355

\bibitem[\protect\citeauthoryear{{Kaaret} \& {Feng}}{{Kaaret} \&
  {Feng}}{2009}]{kaaret09}
{Kaaret}, P.,  \& {Feng}, H. 2009, \apj, 702, 1679

\bibitem[\protect\citeauthoryear{{Kalberla} et~al.}{{Kalberla}
  et~al.}{2005}]{lab}
{Kalberla}, P.~M.~W., {Burton}, W.~B., {Hartmann}, D., {Arnal}, E.~M.,
  {Bajaja}, E., {Morras}, R.,  \& {P{\"o}ppel}, W.~G.~L. 2005, \aap, 440, 775

\bibitem[\protect\citeauthoryear{{Kaur} et~al.}{{Kaur} et~al.}{2011}]{kaur11}
{Kaur}, A., et~al. 2011, ArXiv e-prints

\bibitem[\protect\citeauthoryear{{King} et~al.}{{King} et~al.}{2001}]{king01}
{King}, A.~R., {Davies}, M.~B., {Ward}, M.~J., {Fabbiano}, G.,  \& {Elvis}, M.
  2001, \apjl, 552, L109

\bibitem[\protect\citeauthoryear{{King}, {Kolb}, \& {Szuszkiewicz}}{{King}
  et~al.}{1997}]{king97}
{King}, A.~R., {Kolb}, U.,  \& {Szuszkiewicz}, E. 1997, \apj, 488, 89

\bibitem[\protect\citeauthoryear{{King} \& {Pounds}}{{King} \&
  {Pounds}}{2003}]{kp2003}
{King}, A.~R.,  \& {Pounds}, K.~A. 2003, \mnras, 345, 657

\bibitem[\protect\citeauthoryear{{King} \& {Ritter}}{{King} \&
  {Ritter}}{1998}]{king98}
{King}, A.~R.,  \& {Ritter}, H. 1998, \mnras, 293, L42

\bibitem[\protect\citeauthoryear{{Kolmogorov}}{{Kolmogorov}}{1941}]{kolmogorov41}
{Kolmogorov}, A. 1941, Akademiia Nauk SSSR Doklady, 30, 301

\bibitem[\protect\citeauthoryear{{Kraft}, {Burrows}, \& {Nousek}}{{Kraft}
  et~al.}{1991}]{kraft1991}
{Kraft}, R.~P., {Burrows}, D.~N.,  \& {Nousek}, J.~A. 1991, \apj, 374, 344

\bibitem[\protect\citeauthoryear{{Kuntz} et~al.}{{Kuntz}
  et~al.}{2005}]{kuntz05}
{Kuntz}, K.~D., {Gruendl}, R.~A., {Chu}, Y.-H., {Chen}, C.-H.~R., {Still}, M.,
  {Mukai}, K.,  \& {Mushotzky}, R.~F. 2005, \apjl, 620, L31

\bibitem[\protect\citeauthoryear{{Laird} et~al.}{{Laird}
  et~al.}{2009}]{laird09}
{Laird}, E.~S., et~al. 2009, \apjs, 180, 102

\bibitem[\protect\citeauthoryear{{Lasota}}{{Lasota}}{2001}]{lasota01}
{Lasota}, J.-P. 2001, in Black Holes in Binaries and Galactic Nuclei, ed.
  {L.~Kaper, E.~P.~J.~van den Heuvel, \& P.~A.~Woudt}, 149

\bibitem[\protect\citeauthoryear{{Liu}, {Bregman}, \& {Irwin}}{{Liu}
  et~al.}{2006}]{liu06}
{Liu}, J.-F., {Bregman}, J.~N.,  \& {Irwin}, J. 2006, \apj, 642, 171

\bibitem[\protect\citeauthoryear{{Mapelli}, {Colpi}, \& {Zampieri}}{{Mapelli}
  et~al.}{2009}]{mapelli09}
{Mapelli}, M., {Colpi}, M.,  \& {Zampieri}, L. 2009, \mnras, 395, L71

\bibitem[\protect\citeauthoryear{{Marigo} et~al.}{{Marigo}
  et~al.}{2008}]{marigo08}
{Marigo}, P., {Girardi}, L., {Bressan}, A., {Groenewegen}, M.~A.~T., {Silva},
  L.,  \& {Granato}, G.~L. 2008, \aap, 482, 883

\bibitem[\protect\citeauthoryear{{McClintock} \& {Remillard}}{{McClintock} \&
  {Remillard}}{2006}]{mcclintock06}
{McClintock}, J.~E.,  \& {Remillard}, R.~A. 2006, {Black hole binaries}, ed.
  {Lewin, W.~H.~G.~\& van der Klis, M.} 157

\bibitem[\protect\citeauthoryear{{Mineshige} \& {Wheeler}}{{Mineshige} \&
  {Wheeler}}{1989}]{mineshige89}
{Mineshige}, S.,  \& {Wheeler}, J.~C. 1989, \apj, 343, 241

\bibitem[\protect\citeauthoryear{{Mushotzky}}{{Mushotzky}}{2006}]{mushotzky06}
{Mushotzky}, R. 2006, Advances in Space Research, 38, 2793

\bibitem[\protect\citeauthoryear{{Ohsuga} \& {Mineshige}}{{Ohsuga} \&
  {Mineshige}}{2011}]{om2011}
{Ohsuga}, K.,  \& {Mineshige}, S. 2011, \apj, 736, 2

\bibitem[\protect\citeauthoryear{{Osaki}}{{Osaki}}{1996}]{osaki96}
{Osaki}, Y. 1996, \pasp, 108, 39

\bibitem[\protect\citeauthoryear{{Paczynski}}{{Paczynski}}{1977}]{paczynski77}
{Paczynski}, B. 1977, \apj, 216, 822

\bibitem[\protect\citeauthoryear{{Pakull} \& {Mirioni}}{{Pakull} \&
  {Mirioni}}{2002}]{pakull02}
{Pakull}, M.~W.,  \& {Mirioni}, L. 2002, in New Visions of the X-ray Universe
  in the {\it XMM-Newton} and {\it Chandra} Era, 26-30 November 2001, ESTEC,
  The Netherlands, astro-ph/0202488

\bibitem[\protect\citeauthoryear{{Pilyugin}, {V{\'{\i}}lchez}, \&
  {Thuan}}{{Pilyugin} et~al.}{2006}]{pilyugin2006}
{Pilyugin}, L.~S., {V{\'{\i}}lchez}, J.~M.,  \& {Thuan}, T.~X. 2006, \mnras,
  370, 1928

\bibitem[\protect\citeauthoryear{{Poutanen} et~al.}{{Poutanen}
  et~al.}{2007}]{plfba2007}
{Poutanen}, J., {Lipunova}, G., {Fabrika}, S., {Butkevich}, A.~G.,  \&
  {Abolmasov}, P. 2007, \mnras, 377, 1187

\bibitem[\protect\citeauthoryear{{Prestwich} et~al.}{{Prestwich}
  et~al.}{2011}]{prestwich11}
{Prestwich}, A.~H., {Chandar}, R., {Kuraszkiewicz}, J., {Zezas}, A.,
  {Tsantaki}, M., {Foltz}, R., {Kalogera}, V.,  \& {Linden}, T. 2011, in
  American Astronomical Society Meeting Abstracts \#218, 209.04

\bibitem[\protect\citeauthoryear{{Ptak} et~al.}{{Ptak} et~al.}{2006}]{ptak06}
{Ptak}, A., {Colbert}, E., {van der Marel}, R.~P., {Roye}, E., {Heckman}, T.,
  \& {Towne}, B. 2006, \apjs, 166, 154

\bibitem[\protect\citeauthoryear{{Rappaport}, {Podsiadlowski}, \&
  {Pfahl}}{{Rappaport} et~al.}{2005}]{rapetal05}
{Rappaport}, S.~A., {Podsiadlowski}, P.,  \& {Pfahl}, E. 2005, \mnras, 356, 401

\bibitem[\protect\citeauthoryear{{Roberts}}{{Roberts}}{2007}]{roberts07}
{Roberts}, T.~P. 2007, \apss, 311, 203

\bibitem[\protect\citeauthoryear{{Roberts}, {Levan}, \& {Goad}}{{Roberts}
  et~al.}{2008}]{roberts08}
{Roberts}, T.~P., {Levan}, A.~J.,  \& {Goad}, M.~R. 2008, \mnras, 387, 73

\bibitem[\protect\citeauthoryear{{Romaniello} et~al.}{{Romaniello}
  et~al.}{2002}]{romaniello02}
{Romaniello}, M., {Panagia}, N., {Scuderi}, S.,  \& {Kirshner}, R.~P. 2002,
  \aj, 123, 915

\bibitem[\protect\citeauthoryear{{Ross} \& {Fabian}}{{Ross} \&
  {Fabian}}{2007}]{ross07}
{Ross}, R.~R.,  \& {Fabian}, A.~C. 2007, \mnras, 381, 1697

\bibitem[\protect\citeauthoryear{{Saha} et~al.}{{Saha} et~al.}{2006}]{saha06}
{Saha}, A., {Thim}, F., {Tammann}, G.~A., {Reindl}, B.,  \& {Sandage}, A. 2006,
  \apjs, 165, 108

\bibitem[\protect\citeauthoryear{{Salasnich} et~al.}{{Salasnich}
  et~al.}{2000}]{salasnich00}
{Salasnich}, B., {Girardi}, L., {Weiss}, A.,  \& {Chiosi}, C. 2000, \aap, 361,
  1023

\bibitem[\protect\citeauthoryear{{Shakura} \& {Sunyaev}}{{Shakura} \&
  {Sunyaev}}{1973}]{shakura73}
{Shakura}, N.~I.,  \& {Sunyaev}, R.~A. 1973, \aap, 24, 337

\bibitem[\protect\citeauthoryear{{Sivakoff} et~al.}{{Sivakoff}
  et~al.}{2008}]{sivakoff08}
{Sivakoff}, G.~R., et~al. 2008, \apjl, 677, L27

\bibitem[\protect\citeauthoryear{{Snow} \& {Witt}}{{Snow} \&
  {Witt}}{1996}]{snowwitt1996}
{Snow}, T.~P.,  \& {Witt}, A.~N. 1996, \apjl, 468, L65

\bibitem[\protect\citeauthoryear{{Soria}}{{Soria}}{2007}]{soria07}
{Soria}, R. 2007, \apss, 311, 213

\bibitem[\protect\citeauthoryear{{Soria}}{{Soria}}{2011}]{soria11}
{Soria}, R. 2011, Astronomische Nachrichten, 332, 330

\bibitem[\protect\citeauthoryear{{Soria} et~al.}{{Soria}
  et~al.}{2009}]{soria09}
{Soria}, R., {Risaliti}, G., {Elvis}, M., {Fabbiano}, G., {Bianchi}, S.,  \&
  {Kuncic}, Z. 2009, \apj, 695, 1614

\bibitem[\protect\citeauthoryear{{Soria}, {Wu}, \& {Hunstead}}{{Soria}
  et~al.}{2000}]{soria00}
{Soria}, R., {Wu}, K.,  \& {Hunstead}, R.~W. 2000, \apj, 539, 445

\bibitem[\protect\citeauthoryear{{Stobbart}, {Roberts}, \& {Wilms}}{{Stobbart}
  et~al.}{2006}]{stobbart06}
{Stobbart}, A.-M., {Roberts}, T.~P.,  \& {Wilms}, J. 2006, \mnras, 368, 397

\bibitem[\protect\citeauthoryear{{Strickland} et~al.}{{Strickland}
  et~al.}{2001}]{strickland01}
{Strickland}, D.~K., {Colbert}, E.~J.~M., {Heckman}, T.~M., {Weaver}, K.~A.,
  {Dahlem}, M.,  \& {Stevens}, I.~R. 2001, \apj, 560, 707

\bibitem[\protect\citeauthoryear{{Strickland} \& {Heckman}}{{Strickland} \&
  {Heckman}}{2007}]{strickland07}
{Strickland}, D.~K.,  \& {Heckman}, T.~M. 2007, \apj, 658, 258

\bibitem[\protect\citeauthoryear{{Swartz} et~al.}{{Swartz}
  et~al.}{2004}]{swartz04}
{Swartz}, D.~A., {Ghosh}, K.~K., {Tennant}, A.~F.,  \& {Wu}, K. 2004, \apjs,
  154, 519

\bibitem[\protect\citeauthoryear{{Swartz} et~al.}{{Swartz}
  et~al.}{2011}]{swartz11}
{Swartz}, D.~A., {Soria}, R., {Tennant}, A.~F.,  \& {Yukita}, M. 2011, \apj,
  741, 49

\bibitem[\protect\citeauthoryear{{Tao} et~al.}{{Tao} et~al.}{2011}]{tao11}
{Tao}, L., {Feng}, H., {Grise}, F.,  \& {Kaaret}, P. 2011, ArXiv e-prints

\bibitem[\protect\citeauthoryear{{Titarchuk}}{{Titarchuk}}{1994}]{titarchuk94}
{Titarchuk}, L. 1994, \apj, 434, 570

\bibitem[\protect\citeauthoryear{{Trinchieri}, {Fabbiano}, \&
  {Palumbo}}{{Trinchieri} et~al.}{1985}]{trinchieri85}
{Trinchieri}, G., {Fabbiano}, G.,  \& {Palumbo}, G.~G.~C. 1985, \apj, 290, 96

\bibitem[\protect\citeauthoryear{{van Paradijs} \& {McClintock}}{{van Paradijs}
  \& {McClintock}}{1994}]{van-paradijs94}
{van Paradijs}, J.,  \& {McClintock}, J.~E. 1994, \aap, 290, 133

\bibitem[\protect\citeauthoryear{{Viallet} \& {Hameury}}{{Viallet} \&
  {Hameury}}{2008}]{viallet08}
{Viallet}, M.,  \& {Hameury}, J.-M. 2008, \aap, 489, 699

\bibitem[\protect\citeauthoryear{{Vrtilek} et~al.}{{Vrtilek}
  et~al.}{1990}]{vrtilek90}
{Vrtilek}, S.~D., {Raymond}, J.~C., {Garcia}, M.~R., {Verbunt}, F., {Hasinger},
  G.,  \& {Kurster}, M. 1990, \aap, 235, 162

\bibitem[\protect\citeauthoryear{{Walter} et~al.}{{Walter}
  et~al.}{2008}]{things}
{Walter}, F., {Brinks}, E., {de Blok}, W.~J.~G., {Bigiel}, F., {Kennicutt},
  R.~C., Jr., {Thornley}, M.~D.,  \& {Leroy}, A. 2008, \aj, 136, 2563

\bibitem[\protect\citeauthoryear{{Walton} et~al.}{{Walton}
  et~al.}{2011}]{walton11}
{Walton}, D.~J., {Roberts}, T.~P., {Mateos}, S.,  \& {Heard}, V. 2011, \mnras,
  1147

\bibitem[\protect\citeauthoryear{{Whitehurst}}{{Whitehurst}}{1988}]{whitehurst88}
{Whitehurst}, R. 1988, \mnras, 233, 529

\bibitem[\protect\citeauthoryear{{Winter}, {Mushotzky}, \& {Reynolds}}{{Winter}
  et~al.}{2006}]{winter06}
{Winter}, L.~M., {Mushotzky}, R.~F.,  \& {Reynolds}, C.~S. 2006, \apj, 649, 730

\bibitem[\protect\citeauthoryear{{Zampieri} et~al.}{{Zampieri}
  et~al.}{2011}]{zampieri11}
{Zampieri}, L., {Impiombato}, D., {Falomo}, R., {Gris{\'e}}, F.,  \& {Soria},
  R. 2011, ArXiv e-prints

\bibitem[\protect\citeauthoryear{{Zampieri} \& {Roberts}}{{Zampieri} \&
  {Roberts}}{2009}]{zampieri09}
{Zampieri}, L.,  \& {Roberts}, T.~P. 2009, \mnras, 400, 677

\end{thebibliography}


\begin{deluxetable}{llrrrrrr}
\tablecolumns{7}
\tabletypesize{\scriptsize}
\tablecaption{M83 X-ray Observations
\label{tab:obsid}}
\tablewidth{0pt}
\tablehead{
\colhead{Epoch} &
\colhead{Obsid} &
\colhead{Instrument} &
\colhead{Date} &
\colhead{Exposure} &
\colhead{Flux\tablenotemark{a}} &
\colhead{$L_X$} \\
\colhead{} &
\colhead{} &
\colhead{} &
\colhead{} &
\colhead{(s)} &
\colhead{($10^{-14}$ erg cm$^{-2}$ s$^{-1}$)} &
\colhead{($10^{37}$ erg s$^{-1}$)} }
\startdata
\cutinhead{Swift}
   & 0005605001 & XRT & 2005-01-24 & 8592 & $<$1.75 & $<$4.47 \\
2  & 0031905002 & XRT & 2011-01-03 &  399 & $114\pm33$ & $290\pm83$ \\
3  & 0031905003 & XRT & 2011-01-04 & 1620 & $ 96\pm15$ & $245\pm39$ \\
4  & 0031905004 & XRT & 2011-01-07 & 2213 & $192\pm25$ & $489\pm64$ \\
5  & 0031905005 & XRT & 2011-01-11 & 2140 & $139\pm19$ & $355\pm48$ \\
6  & 0031905006 & XRT & 2011-01-23 & 2896 & $183\pm17$ & $465\pm44$ \\
7  & 0031905007 & XRT & 2011-02-04 & 2938 & $113\pm13$ & $287\pm35$ \\
8  & 0031905008 & XRT & 2011-02-16 & 2882 & $222\pm18$ & $566\pm46$ \\
9  & 0031905009 & XRT & 2011-02-28 & 2863 & $163\pm15$ & $415\pm39$ \\
10\tablenotemark{b} & 0031905010 & XRT & 2011-03-15 & 2285 & $187\pm19$ & $476\pm48$ \\
11\tablenotemark{c} & 0031905011 & XRT & 2011-03-24 & 3258 & $201\pm16$ & $513\pm40$ \\
12 & 0031905012 & XRT & 2011-06-25 & 3240 & $159\pm15$ & $404\pm39$ \\
13 & 0031905013 & XRT & 2011-06-30 & 3146 & $147\pm15$ & $373\pm38$ \\
14\tablenotemark{d} & 0031905014 & XRT & 2011-07-27 & 3588 & $ 94\pm11$ & $239\pm27$ \\
15 & 0031905015 & XRT & 2011-08-24 &  951 & $155\pm33$ & $395\pm84$ \\
16\tablenotemark{e} & 0031905016 & XRT & 2011-08-29 & 2706 & $106\pm15$ & $269\pm39$ \\
17 & 0031905017 & XRT & 2001-09-04 & 4048 & $118\pm12$ & $300\pm31$ \\
\cutinhead{Chandra}
     & 793        & ACIS-S & 2000-04-29 &  50981 & $<0.037$ & $<0.1$ \\
     & 2064       & ACIS-S & 2001-09-04 &   9842 & \ldots & \ldots \\
1A   & 12995      & ACIS-S & 2010-12-23 &  59291 & 120-110 & 360-300 \\
1B   & 13202      & ACIS-S & 2010-12-25 &  98780 & 130-120 & 440-320 \\
10A\tablenotemark{b}  & 12993      & ACIS-S & 2011-03-15 &  49398 & 150-150 & 450-530 \\
10B  & 13241      & ACIS-S & 2011-03-18 &  78963 & 160-150 & 530-410 \\
11A\tablenotemark{c}  & 12994      & ACIS-S & 2011-03-23 & 150058 & 160-150 & 510-420 \\
11B  & 12996      & ACIS-S & 2011-03-29 &  53044 & 150-150 & 530-410 \\
11C  & 13248      & ACIS-S & 2011-04-03 &  54329 & 150-150 & 500-400 \\
16\tablenotemark{e}   & 14332      & ACIS-S & 2011-08-29 &  52381 & 100-100 & 290-290 \\
18   & 12992      & ACIS-S & 2011-09-05 &  66286 & 100-100 & 320-270 \\
19   & 14342      & ACIS-S & 2011-12-28 &  67103 &  80-80  & 230-210 \\
\cutinhead{XMM-Newton}
 & 0110910201 & MOS1 & 2003-01-27 & 21168\tablenotemark{f} & $<0.895$ & $<2.28$ \\
            & & MOS2  &            & 21670 & $<0.879$ & $<2.23$ \\
            & & PN    &            &  9668 & $<0.728$ & $<1.85$ \\
 & 0503230101 & MOS1  & 2008-01-16 & 15946 & not in FOV & \\
            & & MOS2  &            & 16463 & $<1.90$ & $<4.82$ \\
            & & PN    &            &  9490 & not in FOV \\
 & 0552080101 & MOS1  & 2008-08-16 & 25413 & $<1.50$ & $<3.82$ \\
            & & MOS2  &            & 26619 & $<1.65$ & $<4.19$ \\
            & & PN    &            & 15706 & $<0.988$ & $<2.51$ \\
\cutinhead{ROSAT}
 & rh600024a01 & HRI  & 1994-07-30 & 24156 & $<1.0$\tablenotemark{g} & $<2.54$ \\
 & rh600024n00 & HRI  & 1993-01-20 & 23529 & \ldots & \ldots \\
 & rp600188a02 & PSPC & 1993-01-11 & 23350 & \ldots\tablenotemark{h} & \ldots \\
\cutinhead{ASCA}
 & 61016000    & SIS  & 1994-02-12 & 40186 & \ldots\tablenotemark{h} & \ldots \\
\cutinhead{Einstein}
 & I1334S29.XIA& IPC  & 1979-07-31 &  5708 & \ldots\tablenotemark{h} & \ldots \\
 & H1334S29.XIA& HRI  & 1980-01-15 & 20002 & $<13$\tablenotemark{i}  & $<33$ \\
 & H1334S29.XIB& HRI  & 1981-02-13 & 24574 & \ldots & \ldots \\
\enddata
\tablenotetext{a}{In 0.3-10.0 keV.}
\tablenotetext{b}{The \swift\ exposure covered the first part of the \chandra\ exposure;
the \swift\ exposure lasted from 12:17:32 to 22:05:56
while the \chandra\ exposure began at 12:21:40.}
\tablenotetext{c}{The \swift\ exposure covered the end of the \chandra\ exposure;
the \swift\ exposure\ lasted from 11:20:06 to 21:05:57
while the \chandra\ exposure ended at 22:18:33.}
\tablenotetext{d}{00:34:45 to 03:54:36 
This exposure was coincident with the HST exposures in F336E, F438W, F814W.
The HST exposure in F555W occured shortly after the end of the \swift\ exposure.}
\tablenotetext{e}{The \swift\ and \chandra\ exposures were not quite coincident;
the \swift\ exposure lasted from 11:35:00 to 15:13:56 
while the \chandra\ exposure began at 18:41:51.}
\tablenotetext{f}{After soft proton flare cleaning.}
\tablenotetext{g}
{The limiting flux is taken from the $3\sigma$ limit given by \citet{immler99}
who combined the two HRI observations for their analysis.
The luminosity has been calculated from their flux using the \cite{saha06} distance.}
\tablenotetext{h}
{The angular resolution was insufficient to separate the ULX from the nuclear emission.
However, it should be noted that while in the ULX state, 
our source has a flux comparable to that of the nucleus,
and thus would produce a roughly east-west elongation of the nuclear source.
This elongation is not seen for this observation,
suggesting that the source was not in an ultraluminous state.}
\tablenotetext{i}
{The limiting flux is taken from the analysis of \citet{trinchieri85}
who combined the two HRI observations before analysis.
We have used the smallest detected flux from a point-like source in that study for our limit.
The luminosity has been calculated from their flux using the \cite{saha06} distance.
Visual inspection of the images reveal no source at the location of the ULX in either image.}

\end{deluxetable}

\begin{deluxetable}{lcccccccccc}
\tabletypesize{\scriptsize}
\rotate
\tablecolumns{11}
\tablewidth{0pc}
\tablecaption{Best-fit power-law+diskbb Model Parameters\tablenotemark{a}
\label{tab:fit-pldbb}}
\tablehead{
\colhead{Parameter} &  
\multicolumn{10}{c}{Value} \\
\cline{2-11} \\
\colhead{} & 
\colhead{12995} & 
\colhead{13202} & 
\colhead{12993} &
\colhead{13241} & 
\colhead{12994} & 
\colhead{12996} & 
\colhead{13248} &
\colhead{14332} &
\colhead{12992} &
\colhead{14342} }
\startdata
Date & 23 Dec
 & 25 Dec
 & 15 Mar
 & 18 Mar
 & 23 Mar
 & 29 Mar
 & 03 Apr
 & 29 Aug
 & 04 Sep
 & 28 Dec \\
PSFfrac & $0.60^{+0.11}_{-0.26}$
 & $0.58^{+0.11}_{-0.16}$
 & $0.85^{+0.02}_{-0.02}$
 & $0.83^{+0.01}_{-0.01}$
 & $0.84^{+0.01}_{-0.01}$
 & $0.79^{+0.02}_{-0.03}$
 & $0.71^{+0.05}_{-0.07}$
 & $0.71^{+0.15}_{-0.30}$
 & $0.44^{+0.26}_{-0.44}$
 & $0.47^{+0.27}_{-0.47}$ \\
$N_{\rm {H}}$($10^{20}$ cm$^{-2}$) & $4.3^{+2.8}_{-3.0}$
 & $9.2^{+4.0}_{-3.4}$ 
 & $1.1^{+4.2}_{-0.6}$
 & $5.4^{+4.8}_{-1.2}$
 & $3.4^{+0.5}_{-2.7}$
 & $6.1^{+5.8}_{-3.6}$
 & $14.8^{+5.7}_{-4.1}$
 & $3.4^{+2.5}_{-2.5}$
 & $5.3^{+4.5}_{-2.9}$
 & $5.2^{+7.3}_{-4.1}$ \\
$\Gamma$  & $1.98^{+0.16}_{-0.22}$
 & $1.89^{+0.15}_{-0.14}$
 & $1.84^{+0.39}_{-0.36}$
 & $2.15^{+0.18}_{-0.19}$
 & $2.09^{+0.23}_{-0.23}$
 & $2.18^{+0.29}_{-0.24}$
 & $2.26^{+0.19}_{-0.14}$
 & $1.82^{+0.18}_{-0.17}$
 & $1.76^{+0.21}_{-0.16}$
 & $1.60^{+0.17}_{-0.18}$ \\
$N_{\rm {pl}}$($10^{-4}$ phot keV$^{-1}$ s$^{-1}$)\tablenotemark{b} & $2.5^{+0.6}_{-0.4}$
 & $2.5^{+0.3}_{-0.3}$
 & $1.7^{+1.9}_{-0.9}$
 & $2.7^{+2.3}_{-1.4}$
 & $1.9^{+1.0}_{-1.0}$
 & $3.1^{+2.0}_{-1.5}$
 & $4.3^{+0.8}_{-0.5}$
 & $1.8^{+0.3}_{-0.3}$
 & $1.8^{+0.4}_{-0.3}$
 & $1.1^{+0.2}_{-0.3}$ \\
$kT_{\rm{dbb}}$(keV)\tablenotemark{c} & $0.23^{+0.13}_{-0.23}$
 & $0.18^{+0.05}_{-0.03}$
 & $0.38^{+0.03}_{-0.06}$
 & $0.33^{+0.05}_{-0.08}$
 & $0.33^{+0.04}_{-0.02}$
 & $0.32^{+0.06}_{-0.13}$
 & \ldots
 & \ldots
 & $0.25^{+0.10}_{-0.08}$ 
 & $0.26^{+0.15}_{-0.09}$ \\
$K_{\rm{dbb}}$ & $1.1^{+9.1}_{-1.1}$
 & $23.6^{+65.5}_{-18.8}$
 & $2.3^{+2.3}_{-1.1}$
 & $3.9^{+1.7}_{-1.7}$
 & $5.3^{+0.2}_{-1.7}$
 & $3.3^{+2.1}_{-2.0}$
 & \ldots
 & \ldots
 & $1.5^{+15.6}_{-1.4}$
 & $1.2^{+16.8}_{-1.1}$ \\
\tableline\\
$\chi^2$/dof & $1.15$
 & $0.99$
 & $0.94$
 & $1.05$
 & $1.02$
 & $1.15$
 & $1.10$
 & $0.92$
 & $0.94$
 & $0.79$ \\
$\chi^2$ 
 & 250.1
 & 275.3
 & 221.2
 & 296.9
 & 316.4
 & 244.6
 & 232.2
 & 151.2
 & 195.8
 & 148.6 \\
dof 
 & 217
 & 278
 & 236
 & 284
 & 310
 & 212
 & 211
 & 164
 & 209
 & 187 \\
\tableline\\
$C$ (10$^{-1}$ cnt s$^{-1}$)\tablenotemark{de} 
 & $1.43^{+0.02}_{-0.02}$ 
 & $1.50^{+0.01}_{-0.01}$ 
 & $1.93^{+0.02}_{-0.02}$ 
 & $2.07^{+0.02}_{-0.02}$ 
 & $2.13^{+0.01}_{-0.01}$ 
 & $2.01^{+0.01}_{-0.01}$ 
 & $1.90^{+0.02}_{-0.02}$ 
 & $0.98^{+0.01}_{-0.01}$ 
 & $1.11^{+0.01}_{-0.01}$ 
 & $0.83^{+0.01}_{-0.01}$ \\
$f$~($10^{-12}$ erg cm$^{-2}$ s$^{-1}$)\tablenotemark{ef}
 & $1.2^{+0.1}_{-0.1}$
 & $1.3^{+0.1}_{-0.1}$
 & $1.5^{+0.1}_{-0.1}$
 & $1.6^{+0.1}_{-0.1}$
 & $1.6^{+0.1}_{-0.1}$
 & $1.5^{+0.1}_{-0.1}$
 & $1.5^{+0.1}_{-0.1}$
 & $1.0^{+0.1}_{-0.1}$
 & $1.0^{+0.1}_{-0.1}$
 & $0.8^{+0.1}_{-0.1}$ \\
$L$~($10^{39}$ erg s$^{-1}$)\tablenotemark{ef}
 & $3.6^{+0.2}_{-0.2}$
 & $4.4^{+0.5}_{-0.4}$
 & $4.5^{+0.7}_{-0.3}$
 & $5.3^{+1.3}_{-0.7}$
 & $5.1^{+0.5}_{-0.4}$
 & $5.3^{+1.1}_{-0.6}$
 & $\ga 5.0$
 & $2.9^{+0.4}_{-0.4}$
 & $3.2^{+0.5}_{-0.5}$
 & $2.3^{+0.3}_{-0.2}$ \\
$f_{\rm dbb}$\tablenotemark{g} & $\la 3\%$
 & $(17^{+7}_{-4})\%$
 & $(44^{+6}_{-17})\%$
 & $(36^{+12}_{-21})\%$
 & $(51^{+7}_{-10})\%$
 & $(26^{+13}_{-14})\%$
 & \ldots
 & \ldots
 & $(7^{+2}_{-6})\%$
 & $(9^{+6}_{-4})\%$ \\
\enddata
\tablenotetext{a}
 {The complete spectral model fitted was 
  pileup $\times$ tbabs $_{Gal}\times$ tbabs $_i\times$ (power-law+diskbb).
  All uncertainties are for the 90\% confidence interval.}
\tablenotetext{b}
 {At 1 keV.}
\tablenotetext{c}
 {$K_{\rm{dbb}}=\left[r_{\rm in}({\rm km})/d(10{\rm ~kpc})\right]^2 
  \times\cos\theta$ where $r_{\rm in}$ is the apparent inner-disk radius
  and $\theta$ the viewing angle.}
\tablenotetext{d}{Before pile-up correction.}
\tablenotetext{e}{In 0.3-10.0 keV.}
\tablenotetext{f}{After pile-up correction.}
\tablenotetext{g}
  {Fraction of unabsorbed X-ray luminosity in the diskbb component.}
\end{deluxetable}

\begin{deluxetable}{lcccccccccc}
\tabletypesize{\scriptsize}
\rotate
\tablecolumns{11}
\tablewidth{0pc}
\tablecaption{Best-fit comptt Model Parameters\tablenotemark{a}
\label{tab:fit-comptt}}
\tablehead{
\colhead{Parameter} &  
\multicolumn{10}{c}{Value} \\
\cline{2-11} \\
\colhead{} & 
\colhead{12995} &
\colhead{13202} &
\colhead{12993} &
\colhead{13241} &
\colhead{12994} &
\colhead{12996} &
\colhead{13248} &
\colhead{14332} &
\colhead{12992} &
\colhead{14342} }
\startdata
Date & 23 Dec
 & 25 Dec
 & 15 Mar
 & 18 Mar
 & 23 Mar
 & 29 Mar
 & 03 Apr
 & 29 Aug
 & 04 Sep
 & 28 Dec \\
PSFfrac & $0.59^{+0.11}_{-0.17}$
 & $0.64^{+0.07}_{-0.09}$
 & $0.79^{+0.04}_{-0.05}$
 & $0.77^{+0.03}_{-0.03}$
 & $0.79^{+0.02}_{-0.02}$
 & $0.73^{+0.04}_{-0.05}$
 & $0.75^{+0.05}_{-0.06}$
 & $0.68^{+0.15}_{-0.31}$
 & $0.47^{+0.20}_{-0.47}$ 
 & $0.42^{+0.29}_{-0.42}$ \\
$N_{\rm {H}}$($10^{20}$ cm$^{-2}$) & $<1.6$
 & $<1.0$
 & $7.2^{+2.5}_{-7.2}$
 & $<8.0$
 & $<1.2$
 & $<4.9$
 & $4.0^{+3.0}_{-3.4}$
 & $<4.5$
 & $<3.5$
 & $<3.5$ \\
$kT_{0}$ (keV)\tablenotemark{b}  & $0.11^{+0.02}_{-0.01}$
 & $0.12^{+0.01}_{-0.01}$
 & $<0.12$
 & $0.15^{+0.01}_{-0.06}$
 & $0.14^{+0.01}_{-0.01}$
 & $0.14^{+0.01}_{-0.04}$
 & $0.11^{+0.02}_{-0.02}$
 & $<0.14$
 & $0.13^{+0.02}_{-0.03}$
 & $0.13^{+0.02}_{-0.04}$ \\
$kT_{e}$ (keV)\tablenotemark{c}  & $1.9^{+1.2}_{-0.2}$
 & $>2.0$
 & $>6.5$
 & $>1.8$
 & $>1.9$
 & $>1.9$
 & $>2.6$
 & $>1.8$
 & $>1.8$ 
 & $>2.2$ \\
$\tau$\tablenotemark{d} & $7.4^{+0.9}_{-0.8}$
 & $<8.3$
 & $1.8^{+5.3}_{-0.5}$
 & $5.0^{+2.6}_{-2.7}$
 & $4.9^{+1.4}_{-0.7}$
 & $5.0^{+2.3}_{-1.1}$
 & $0.7^{+3.9}_{-0.3}$
 & $3.2^{+16.4}_{-0.6}$
 & $7.6^{+0.7}_{-1.5}$
 & $5.9^{+1.7}_{-1.6}$ \\
$K_{\rm{c}}$\tablenotemark{e} & $4.5^{+1.2}_{-1.9}$
 & $3.5^{+1.2}_{-\ast}$
 & $2.8^{+\ast}_{-\ast}$
 & $5.6^{+6.0}_{-\ast}$
 & $6.9^{+0.7}_{-\ast}$
 & $4.9^{+4.3}_{-\ast}$
 & $0.3^{+1.0}_{-\ast}$
 & $0.4^{*}_{*}$\tablenotemark{f}
 & $2.7^{+4.2}_{-\ast}$
 & $1.0^{+4.2}_{-\ast}$ \\
\tableline\\[-10pt]
$\chi^2$/dof & $1.10$
 & $0.95$
 & $0.97$
 & $1.04$
 & $1.05$
 & $1.15$
 & $1.10$
 & $0.93$
 & $0.93$
 & $0.81$ \\
$\chi^2$
 & 239.2
 & 263.5
 & 229.5
 & 294.5
 & 325.2
 & 242.7
 & 231.3
 & 151.4
 & 194.8
 & 151.1 \\
dof
 & 217
 & 278
 & 236
 & 284
 & 310
 & 212
 & 211
 & 162
 & 209
 & 187 \\
\tableline\\
$C$~($10^{-1}$ cnt s$^{-1}$)\tablenotemark{gh} & $1.43 \pm 0.02$
 & $1.50^{+0.01}_{-0.01}$ 
 & $1.93^{+0.02}_{-0.02}$
 & $2.07^{+0.02}_{-0.02}$
 & $2.13^{+0.01}_{-0.01}$
 & $2.01^{+0.01}_{-0.01}$
 & $1.90^{+0.02}_{-0.02}$
 & $0.98^{+0.01}_{-0.01}$
 & $1.11^{+0.01}_{-0.01}$
 & $0.83^{+0.01}_{-0.01}$ \\
$f$~(erg cm$^{-2}$ s$^{-1}$)\tablenotemark{hi} & $1.1^{+0.1}_{-0.1}$
 & $1.2^{+0.1}_{-0.1}$
 & $1.5^{+0.1}_{-0.1}$
 & $1.5^{+0.1}_{-0.1}$
 & $1.5^{+0.1}_{-0.1}$
 & $1.5^{+0.1}_{-0.1}$
 & $1.5^{+0.1}_{-0.1}$
 & $1.0^{+0.1}_{-0.1}$
 & $1.0^{+0.1}_{-0.1}$
 & $0.8^{+0.1}_{-0.1}$ \\
$L$~($10^{39}$ erg s$^{-1}$)\tablenotemark{hj} & $3.0^{+0.1}_{-0.1}$
 & $3.2^{+0.1}_{-0.1}$
 & $5.3^{+0.5}_{-1.2}$
 & $4.1^{+1.4}_{-0.1}$
 & $4.2^{+0.2}_{-0.2}$
 & $4.1^{+0.6}_{-0.2}$
 & $4.0^{+0.9}_{-0.1}$
 & $2.9^{+0.4}_{-0.4}$
 & $2.7^{+0.2}_{-0.2}$
 & $2.1^{+0.2}_{-0.2}$ \\
\enddata
\tablenotetext{a}
 {The complete spectral model fitted was
  pileup $\times$ tbabs $_{Gal}\times$ tbabs $_i\times$ (comptt).
  All uncertainties are for the 90\% confidence interval.}
\tablenotetext{b}{Seed photon temperature.}
\tablenotetext{c}{Plasma temperature.}
\tablenotetext{d}{Scattering optical depth.}
\tablenotetext{e}{Normalization in $10^{-4}$ XSPEC units (Titarchuk 1994).}
\tablenotetext{f}{Uncertainty does not converge as the temperature is unconstrained.}
\tablenotetext{g}{Before pile-up correction.}
\tablenotetext{h}{In 0.3-10.0 keV.}
\tablenotetext{i}{After pile-up correction.}
\tablenotetext{j}{Unabsorbed luminosity.}
\end{deluxetable}

\begin{deluxetable}{lcccccccccc}
\tabletypesize{\scriptsize}
\rotate
\tablecolumns{11}
\tablewidth{0pc}
\tablecaption{Best-fit Power-Law Over the $2$--$10$ keV Band
\label{tab:fit-pow}}
\tablehead{
\colhead{Parameter} &  
\multicolumn{10}{c}{Value} \\
\cline{2-11} \\
\colhead{} &
\colhead{12995} &
\colhead{13202} &
\colhead{12993} &
\colhead{13241} &
\colhead{12994} &
\colhead{12996} &
\colhead{13248} &
\colhead{14332} &
\colhead{12992} &
\colhead{14342} }
\startdata
$\Gamma$ & $2.05^{+0.10}_{-0.13}$
 & $1.97^{+0.13}_{-0.06}$
 & $2.50^{+0.10}_{-0.12}$
 & $2.61^{+0.08}_{-0.09}$
 & $2.73^{+0.06}_{-0.06}$
 & $2.45^{+0.14}_{-0.08}$
 & $2.36^{+0.08}_{-0.13}$
 & $1.84^{+0.14}_{-0.13}$
 & $1.85^{+0.16}_{-0.10}$
 & $1.61^{+0.33}_{-0.13}$ \\
 $\chi^2_{\nu}$& $1.06$
 & $0.90$
 & $0.85$
 & $1.12$
 & $0.97$
 & $1.13$
 & $1.05$
 & $0.91$
 & $0.84$
 & $0.81$ \\
$\chi^2$ 
 & 114.58
 & 151.62
 & 108.22
 & 195.85
 & 194.40
 & 117.19
 & 109.25
 & 56.4
 & 86.6
 & 66.1 \\
dof
 & 108
 & 168
 & 128
 & 175
 & 201
 & 104
 & 104
 & 62
 & 103
 & 82 \\
\tableline\\
$\Gamma_1$ & $1.99^{+0.11}_{-0.17}$
 & $1.94^{+0.10}_{-0.09}$
 & NA\tablenotemark{a}
 & NA
 & NA
 & NA
 & NA
 & NA
 & NA
 & NA \\
$E_{\rm br}$~(keV) & $5.3^{+3.2}_{-3.7}$
 & $6.0^{+0.7}_{-2.0}$
 & NA
 & NA
 & NA
 & NA
 & NA
 & NA
 & NA
 & NA \\
$\Gamma_2$ & $3.14^{+4.48}_{-1.29}$
 & $4.56^{+\ast}_{-2.41}$
 & NA
 & NA
 & NA
 & NA
 & NA
 & NA
 & NA
 & NA \\
$\chi^2_{\nu}$& $1.06$
 & $0.88$
 & NA
 & NA
 & NA
 & NA
 & NA
 & NA
 & NA
 & NA \\
$\chi^2$ 
 & 111.92
 & 146.74
 & NA
 & NA
 & NA
 & NA
 & NA
 & NA
 & NA
 & NA \\
dof
& 106
& 166
 & NA
 & NA
 & NA
 & NA
 & NA
 & NA
 & NA
 & NA \\
F test & $P = 0.288$
 & $P = 0.066$
 & NA
 & NA
 & NA
 & NA
 & NA
 & NA
 & NA
 & NA \\
\enddata
\tablenotetext{a}
  {Broken power-law models with the constraint that
   $\Gamma_2 > \Gamma_1$ do not provide any improvement 
   to the $\chi^2_{\nu}$ value.}
\end{deluxetable}


\begin{deluxetable}{lccccccccr}
\tabletypesize{\scriptsize}
\rotate
\tablecolumns{5}
\tablewidth{0pc}
\tablecaption{Observations of  the Optical Counterpart to the M83 ULX
\label{tab:opti}}
\tablehead{
\colhead{Date}  & \colhead{Telescope} &\colhead{Instrument} & \colhead{Exposure} & \multicolumn{3}{c}{Filter}  
& \colhead{Magnitude\tablenotemark{a}} &\colhead{A($\lambda$)\tablenotemark{b}} &\colhead{Unabsorbed}  \\
\cline{5-7}
\colhead{} & \colhead{}&\colhead{} & \colhead{(s)} & \colhead{Band} & \colhead{$\lambda_c$ (\AA)} & \colhead{$\Delta\lambda$ (\AA)}& \colhead{} &\colhead{(mag)} &\colhead{Flux (Jy)\tablenotemark{c}}
}
\startdata
2009 Apr 26 & Magellan I  & IMACS & $ 7 \times 600 $ & [O\thinspace III]  &5007 & 50 &  \\
2009 Apr 26 & Magellan I  & IMACS & $ 7 \times 200$ & Green Cont'm & 5316 & 161 & $m_{5150} \gtrsim 25.0$ \\
2009 Apr 26 & Magellan I  & IMACS & $ 7 \times 200$ & Red Cont'm & 6815 & 216 &  \\
\tableline
2009 Aug 9 & {\it HST} & WFC3  & $3 \times 630$ & {\it U} (F336W)  & 3355 & 511 &  $m_{3355} =28.1\pm 0.8$\tablenotemark{c}& 0.48 & $3.2\times 10^{-8}$\\
2009 Aug 9 & {\it HST} & WFC3  & $3 \times 640$ & {\it B} (F438W)  & 4326 & 618 &  $m_{4326} =27.5\pm 0.6$\tablenotemark{c} &0.40 & $5.4\times 10^{-8}$\\
2009 Aug 9 & {\it HST} & WFC3  & $3 \times 401 $ & {\it V} (F555W)  &  5308 & 1562 & $m_{5308} = 26.4 \pm 0.3$\tablenotemark{c} & 0.30 & $1.3\times 10^{-7}$\\ 
2009 Aug 9  & {\it HST} & WFC3  & $3 \times 401 $ & {\it I} (F814W)  &  8030 & 1536 & $m_{8030} =24.4\pm 0.3$\tablenotemark{c} & 0.18 & $7.7\times 10^{-7}$\\ 
\tableline
\tableline
2011 Apr 8 & Gemini-S & GMOS & $4 \times 600$  & $u^\prime$ & 3650 & 490 & $m_{3650} = 23.4 \pm 0.3$ & 0.47 & $2.4\times 10^{-6}$\\
2011 Apr 8 & Gemini-S & GMOS &$4 \times 100 $ & $g^\prime$ & 4780 & 1540 & $m_{4780} = 23.8 \pm 0.2$& 0.35 & $1.6\times 10^{-6}$ \\
2011 Apr 8 & Gemini-S & GMOS & $4 \times 150$ & $r^\prime$ & 6340 & 1440 & $m_{6340} = 23.8 \pm0.3 $& 0.25 & $1.4\times 10^{-6}$ \\
\tableline
2011 July 27  & {\it HST} & WFC3  & $3 \times 608 $ & {\it U} (F336W)  &  3355 & 511 & $m_{3355}=23.18\pm0.08$ & 0.48 & $2.0\times 10^{-6}$\\ 
2011 July 27  & {\it HST} & WFC3  & $3 \times 330 $ & {\it B} (F438W)  &  4326 & 618 & $m_{4326} =23.71\pm0.06$ & 0.40 & $1.7\times 10^{-6}$\\ 
2011 July 27  & {\it HST} & WFC3  & $3 \times 245 $ & {\it V} (F555W)  &  5308 & 1562 & $m_{5308} =23.73\pm0.05$ & 0.30 & $1.6\times 10^{-6}$\\ 
2011 July 27  & {\it HST} & WFC3  & $3 \times 487 $ & {\it I} (F814W)  &  8030 & 1536 & $m_{8030}=23.87\pm0.06$ & 0.18 & $1.7\times 10^{-6}$\\ 

\enddata
\tablenotetext{a}{All magnitudes are observed AB magnitudes at the central wavelength.
For WFC3, these use the nominal WFC3 zero points given at http://www.stsci.edu/hst/wfc3/phot\_zp\_lbn.}
\tablenotetext{b}{Assuming $A_V = 0.30$ and using reddening corrections from \citet{romaniello02} for the HST filters and  from  \citet{cardelli89} at the central wavelengths for the Gemini filters.}
\tablenotetext{c}{Values obtained from the 2009 {\it HST} observations correspond to the excess flux at the location of the ULX
and are upper limits to the flux of the pre-outburst donor star.}
\end{deluxetable}

\clearpage

\begin{figure*}
\plotone{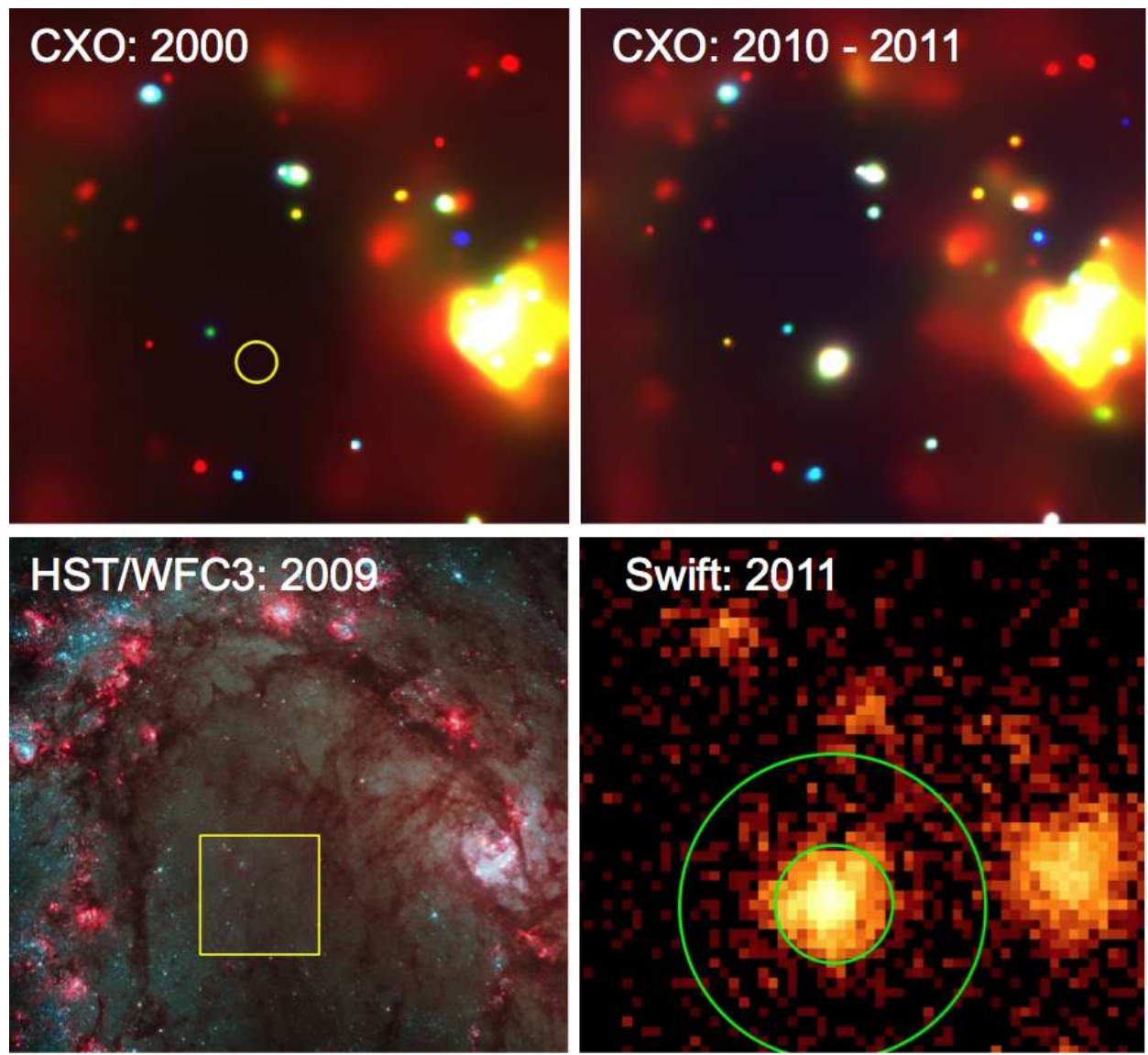}
\caption{
{\bf Upper Left:} X-ray image of the inner region of M83,
from the \chandra\ observation in 2000.
Color coding: red, $0.3$--$1.0$ keV; green, $1.0$--$2.0$ keV; blue, $2.0$--$7.0$ keV.
The yellow circle encloses the position of the ULX.
{\bf Upper Right:} X-ray image of the same region from our 2010--2011
\chandra\ observations, showing the newly discovered ULX.
Color coding is as in the upper left panel.
{\bf Lower Left:} Optical color composite image of approximately the
same region from the 2009 \hst /WFC3 Early Release Science data.  
The diffuse red emission is from H$\alpha$ 
and U, V, and I bands are shown as blue, green, and red.
The $30\arcsec \times 30\arcsec$ yellow box is shown in greater detail in Figure~\ref{fig:gem_opt_3}.
{\bf Lower Right:} The co-added \swift\ image in the $0.3$--$10.0$ keV band.
The source and background regions used for the photometry are shown.
Each panel is just over $2\arcmin$ on a side
and the physical size of each panel is $\approx 2.7$ kpc.  
North is up, East is left.
\label{fig:find_chart}}
\end{figure*}

\begin{figure}
\plotone{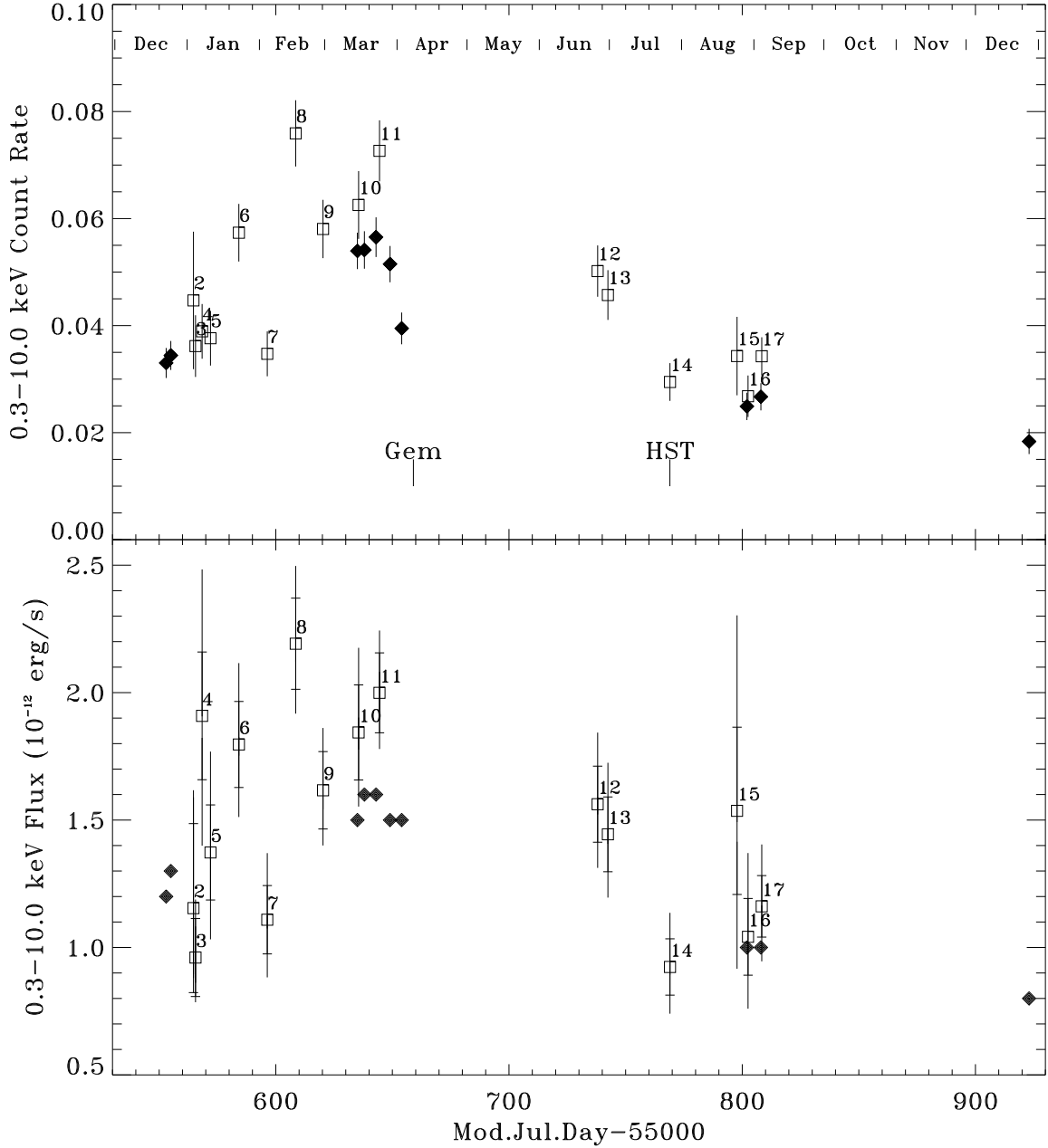}
\vspace{0.3in}
\caption{
{\bf Top:} The \swift\ $0.3$--$10.0$ keV count rate as a function of time.
The open boxes are the \swift\ points labeled by their epoch.
The filled diamonds are the \chandra\ {\it powerlaw+diskbb} spectral fits converted to \swift\ count rates.
The dates of the Gemini and \hst\ observations are also marked.
{\bf Bottom:}  The  flux as a function of time.
The open boxes are the \swift\ count rates converted to fluxes as described in the text.
The filled diamonds are the \chandra\ fluxes derived from the {\it powerlaw+diskbb} spectral fits.
\label{fig:lc}}
\end{figure}

\begin{figure}
\plotone{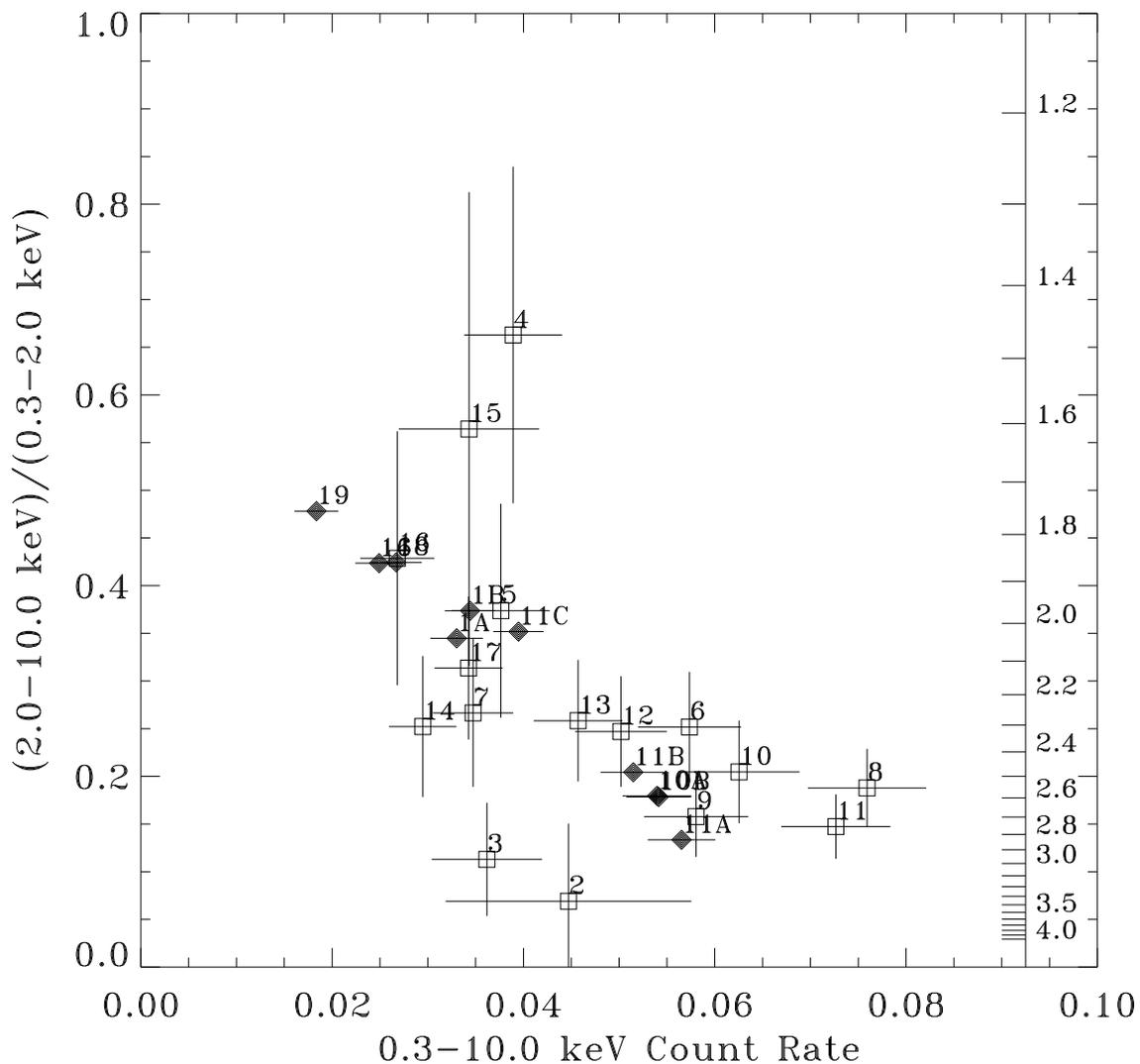}
\caption{
The $2.0$--$10.0$ keV/$0.3$--$2.0$ keV hardness ratio
as a function of the $0.3$--$10.0$ keV band count rate.
The scale on the right is the equivalent power law index,
assuming a simple absorbed power law with an N(H) of
$1.2\times10^{21}$ cm$^{-2}$.
The points are labeled by their epoch.
The open boxes are the \swift\ data points labeled by their epoch.
The filled diamonds are the \chandra\ {\it powerlaw+diskbb} spectral fits
converted to the \swift\ count rates and hardness ratios.
\label{fig:cr_hr}}
\end{figure}

\begin{figure*}
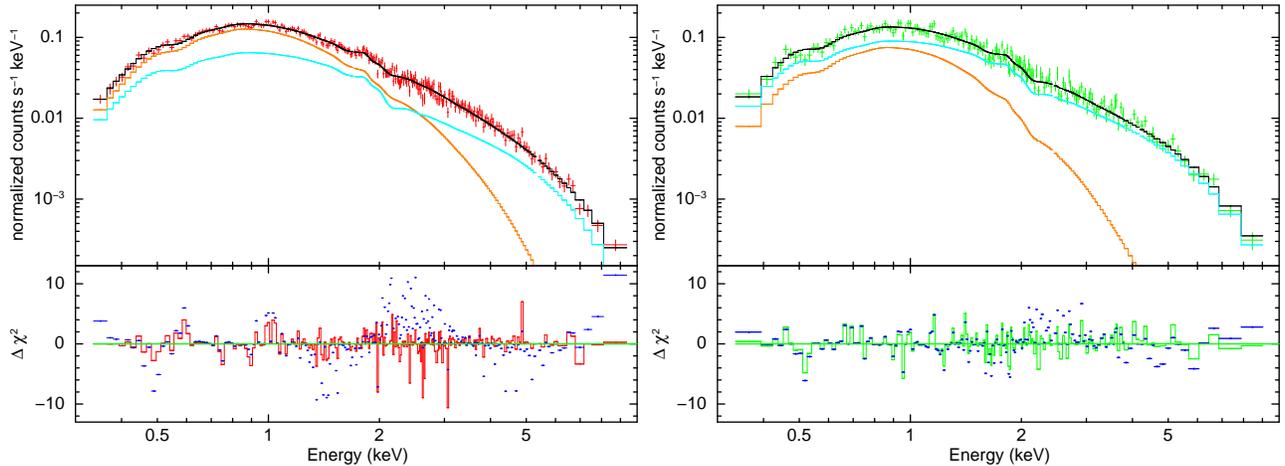

\centerline{\includegraphics[angle=-90,scale=0.35]{fig4a.eps}
\includegraphics[angle=-90,scale=0.35]{fig4b.eps}}
\caption{{\bf Left:}
The 23 March 2011 {\it  Chandra}/ACIS spectrum 
fitted with an absorbed {\it diskbb} plus {\it powerlaw} model, 
convolved with a {\it pileup} model.
The data points (in red) are binned to a signal-to-noise ratio $>7$
for display purposes only. 
{\bf Top panel:} The separate contributions of the two model components 
({\it diskbb} in orange, {\it powerlaw} in cyan).
{\bf Bottom panel:} The $\chi^2$ residuals for the best-fitting model
with pile-up correction (red stepped line), 
as well as without pile-up correction (blue data points). 
See Table 2 for the best-fitting parameters.
{\bf Right:} As in the left panel but
for the 29 March 2011 spectrum.
The data points (in green) are binned to a signal-to-noise ratio $>5$
for display purpose only.
See Table 2 for the best-fitting parameters.
Note the change in relative strength of the fit components.}
\label{fig:chandra_spec}
\end{figure*}

\begin{figure*}
\plotone{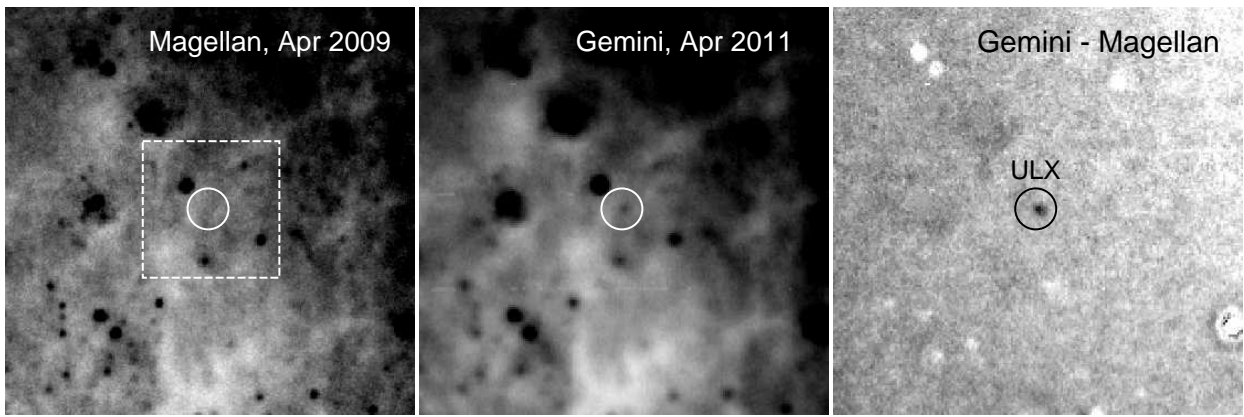}
\caption{
Images showing a 30\arcsec\ square region centered on the position of the ULX 
(the 3\arcsec\ diameter circle) from 
{\bf Left:} Magellan IMACS, 
([O~III] plus narrow green continuum) from April 2009,  
{\bf Center:} Gemini-S GMOS, ({\it g}) from April 2011, and 
{\bf Right:} the difference image between the Gemini image 
and the Magellan one after PSF-matching and scaling.
North is up and east to the left; 
the dashed box indicates the 10\arcsec\ field shown in Fig.~\ref{fig:wfc3}.}
\label{fig:gem_opt_3}
\end{figure*}

\begin{figure*}
\plotone{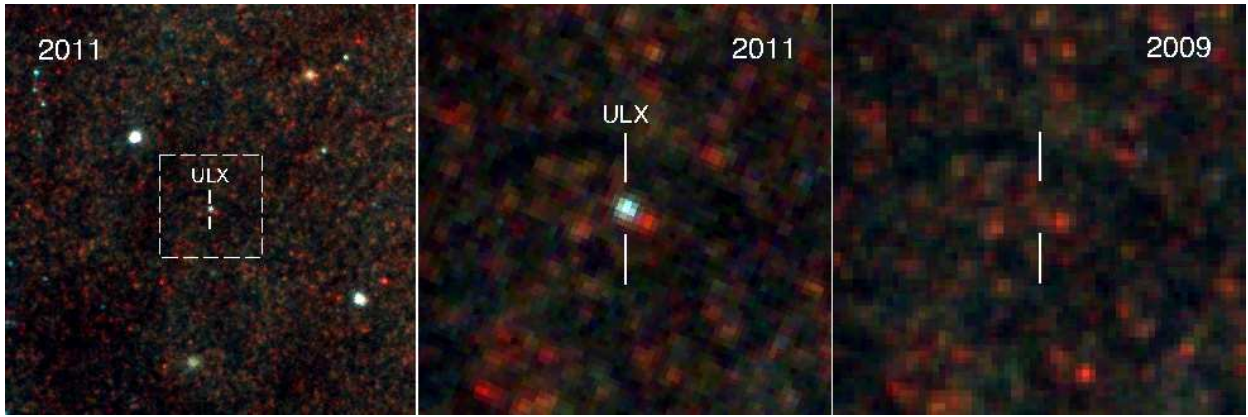}
\caption{
\hst\ WFC3 true-color images of the region around the ULX.
In all panels, the colors are: blue, for the F438W filter; 
green, for F555W; and red, for F814W. North is up and East to the left.
{\bf Left:} Image from July 2011, during the X-ray luminous state; 
the ULX counterpart appears distinctly blue.
The field size is 10\arcsec\ square.
This panel can be compared to the ground-based images in the previous figure.
The dashed box shows the region covered by the next two panels.
{\bf Middle:} A $2\farcs5$ by $2\farcs5$ detail of the July 2011 image.
{\bf Right:} Image from August 2009, at the same scale as the previous panel.
The bright blue ``star'' is not present; 
there are many faint red stars near the ULX position in the
2009 image, but none is exactly coincident.
\label{fig:wfc3}}
\end{figure*}

\begin{figure*}
\includegraphics[angle=-90,scale=.6]{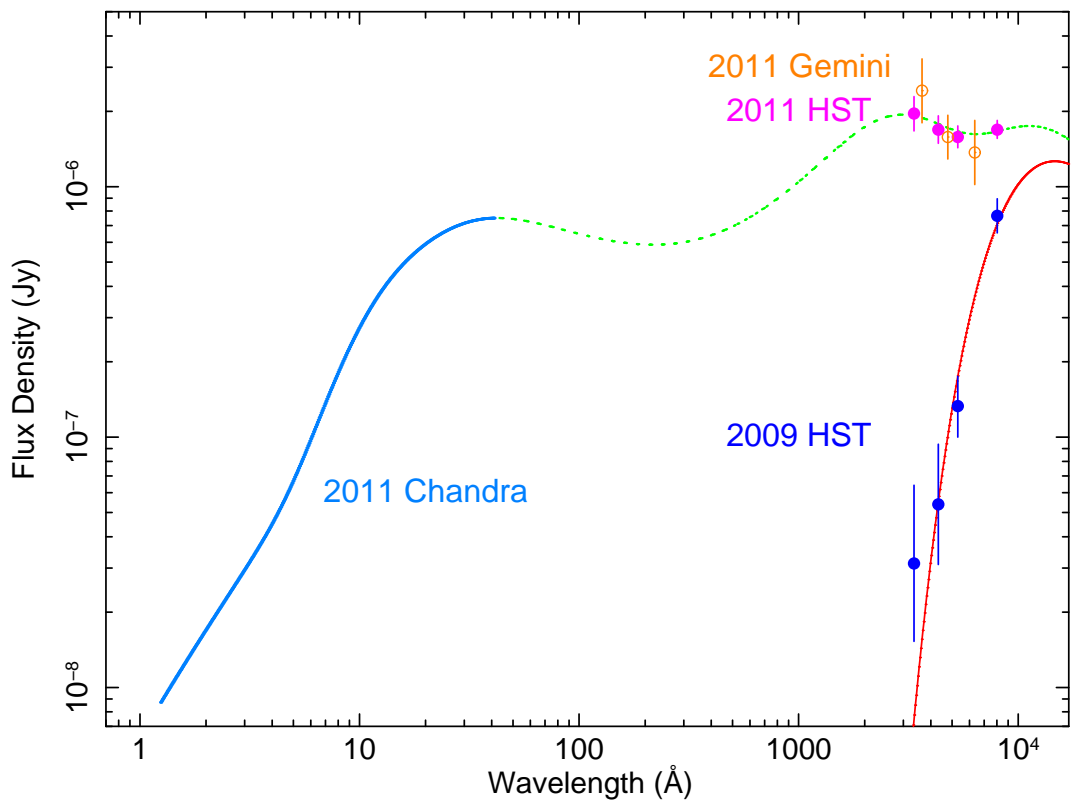}
\caption{
Model of the multi-wavelength spectral energy distribution (SED) 
in outburst and quiescence.
The {\bf blue} curve at short wavelengths is the $0.3$--$10$ keV 
best fit {\it diskir} model for the 23 March 2011 \chandra\ data,
which were chosen for this exercise as they are from the longest observation
with the strongest constraints on the accretion disk parameters.
The {\bf dark blue} data points are the pre-outburst \hst\ measurements 
from 9 August 2009. 
These are actually upper limits for the flux from any actual counterpart.
The {\bf red} curve is the best fit 
assuming that the upper limits to the light from the pre-outburst star actually correspond to the light from that star.
The {\bf magenta} data points are from the 27 July 2011 \hst\ measurements;
the {\bf orange} data points are from the 8 April 2011 Gemini measurements;
and the {\bf green} dotted line is the modeled emission from an irradiated disk
with the addition of the pre-outburst blackbody.
All data and models have been corrected for absorption/extinction.
}
\label{fig:opt_xray}
\end{figure*}


\end{document}